Network Evolution and National Interests: Global Scientific Reorganization and the Rise of Scientific Nationalism

Caroline S. Wagner[1]; Xiaojing Cai[2]

**Abstract**

The global network of scientific cooperation has undergone fundamental restructuring over the past two decades, with significant implications for geopolitical relations and science policy, and, in part, shaped by them. China's integration into this network has redistributed positions of power and influence in ways that challenge conventional zero-sum narratives around national scientific competition, sovereignty, and security. Drawing on structural holes theory (Burt, 1992; 2004) and the Bianconi-Barabási fitness model (2001), we propose that China's entry into the global network over the past 20 years accelerated a structural reorganization already underway through network maturation. As China's rising scientific attractiveness drew direct collaborative ties that bypassed U.S. intermediation, the structural holes spanned by the United States were filled by alternative pathways. The network analysis reveals an asymmetric pattern: U.S. topological betweenness centrality declined substantially (roughly 80%) while weighted betweenness centrality, which incorporates collaboration volume, remained comparatively stable. This suggests for the U.S. distinct losses of brokerage capital (the structural advantage of bridging disconnected communities and controlling the flow of information between them) and closure capital (the advantage of deep bilateral relationships that sustain collaborative productivity independent of brokerage function). Granger causality tests show that China's early network participation predicted subsequent structural changes across multiple metrics and fields, consistent with its role as a high-fitness entrant. Field-level analysis consistent across six major domains confirms the drop in U.S. centrality regardless of the position of China or other nations.

[1] The Ohio State University, Columbus, OH USA; 0000-0002-1724-8489
[2] Yangzhou University, Jiangsu, China; 0000-0001-7346-6029

**The Growth of a Global Network of Scientific Communication**

A notable feature of publicly funded research over the past three decades has been the shift from nationally organized science to broad participation in an internationally connected communications network of researchers. International collaborations, represented by co-authorships in scholarly publications, grew from 4% in the Web of Science in 1980 to 27% by 2020 (Gui et al., 2025; Docampo & Bessoule, 2019; Leydesdorff & Wagner, 2008; Howells, 1990). Although not necessarily more novel evidence shows that international collaborations attract more citations than domestic work, producing both higher output and greater impact (Frenken, Ponds & Van Oort, 2010; Wagner et al., 2019; Jefferson et al., 2024). The more elite the researcher, the more likely they are to work at the international level (Tang, 2013). In scientifically advanced countries, our research shows all the growth has been at the international level.

This shift is more than a quantitative increase in cross-border collaboration; it represents a qualitative transformation in the organizational logic of science. The global network grew significantly after 1980, arguably as a self-organizing system with its own structural dynamics — ones that exist, in important respects, orthogonal to the national systems that fund, train, and support its participants (Adams & Szomszor, 2022; Pan et al., 2012; Rosvall & Bergstrom, 2007; Börner et al., 2003).

The theoretical basis for this claim was established in earlier work testing whether international collaboration could be explained by conventional structural, institutional, or policy factors — such as centre-periphery dynamics, the growth of megascience, geographic proximity, and the expansion of information and communications technologies (Wagner et al., 2015; Wagner & Leydesdorff, 2005). None of these factors, individually or in combination, proved adequate to explain the observed growth of international co-authorship. What the data instead revealed was a self-organizing system evolving by a mechanism of preferential attachment: researchers seek out partners who are already well-connected, offering access to resources, reputation, and research opportunities (Jeong et al., 2003). The network grows not through institutional direction but through the aggregated self-interested choices of individual scientists operating within a competitive reward structure (Milojevic, 2014). This finding situates the global scientific network firmly within the class of complex adaptive systems identified by Barabási and Albert (1998), Leydesdorff (2001), and Dijk (2005), in which order emerges from local interactions rather than centralized coordination.

Our analysis examines the global network of scientific collaboration over 20 years, revealing a network that undergoes recurring structural shifts: the number of detected communities rises and falls across the period, reflecting cycles of fragmentation and consolidation as new actors enter and collaborative patterns reorganize in response. The fluctuation in community numbers is not in itself the significant finding — what matters is the character of each successive reorganization. Rather than simply oscillating, the network produces communities that are increasingly well-defined and internally cohesive with each cycle, suggesting that the reorganization is cumulative rather than merely turbulent. Such behavior is characteristic of complex adaptive systems in which periods of apparent instability are the mechanism through which more advantageous organizational structures emerge.

**Theoretical Framework: Structural Position, Brokerage, and the Reorganization of Scientific Networks**

The relationship between geopolitical context and the network structure appears neither direct nor determinate and it is understudied and undertheorized. Our earlier work with Leydesdorff found that the global network grows "denser but not more clustered," and that "power relationships within the network do not reproduce those of the political system" — suggesting that network dynamics generate structural outcomes that diverge systematically from the international political order (Wagner & Leydesdorff, 2015; Wagner & Leydesdorff, 2005). This suggests that scientific communications operate and evolve according to network dynamics rather than political entitied. Lee et al. (2020), supported by Fry et al. (2021), demonstrated this empirically in an acute case: US-China scientific collaboration on COVID-19 research actually increased amid intensifying political rhetoric, suggesting the network's collaborative logic can operate with considerable autonomy from its political environment. Glänzel et al. (2004) confirmed that co-authorship analysis provides a reliable bibliometric instrument for tracking these dynamics. The evidence supports neither a purely structural account in which the network operates free of geopolitical influence, nor a purely political account in which collaboration simply mirrors the international order. The relationship is complex and context-dependent.

The data suggests significant changes in the network, especially with the entrance of China. What the literature has not yet provided is a network-level theoretical account of the influence of changing network dynamics on science policy. This paper offers such an account, drawing on structural holes theory to interpret the structural transformation our data document, drawing upon the rise of China, the shifting communications patterns, and what this has meant for the role of United States.

Since at least 1980, the United States centered the global network. As the world's leading scientific power in the post-Cold War period, it attracted disproportionate early connections, and each new connection reinforced its centrality through cumulative advantage. By 2000, the U.S. appeared in the vast majority of the shortest paths connecting any two countries in the network—a measure of its near-total intermediary role. Preferential attachment explains how this position was built. It does not, however, explain what that position was worth, why it eroded, or what the erosion means for science policy. For those questions, we turn to a different framework.

Burt's structural holes theory (1992; 2004) provides this framework. Its central insight is that the competitive advantage of a network position derives not from the number of connections a node holds but from their pattern — specifically, from whether those connections link to otherwise disconnected communities or to nodes that are already connected to one another. A structural hole is the absence of a direct connection between two nodes that share a common contact. A node spanning such a hole sits at the junction of separate communities and gains three distinct advantages: broader and earlier access to information circulating in non-overlapping scientific worlds; a degree of control over what each community knows about the other; and what Burt (2010) calls 'structural vision'—the ability to see across the distributed landscape of a system in ways unavailable to nodes embedded within a single cluster. In a scientific network, these advantages are simultaneously social and epistemic: the broker controls not only relationships but the flow of knowledge itself. These roles provided the U.S. science system with great advantages, insights into scientific advances around the world, ready access to top talent, and a diffusion mechanism for U.S. knowledge.

The literature distinguishes two complementary network logics that generate different forms of advantage to network participants (Burt, 2004; Coleman, 1988). Brokerage, defined as vspanning holes between disconnected communities produces information and vision advantages. It provides the advantage of deep, voluminous bilateral relationships that sustain collaborative productivity independent of brokerage function. Closure is defined as dense, cohesive ties within a community, producing trust, shared norms, and the capacity for sustained collective effort. For the central hub, closure provides the advantage of deep, voluminous bilateral relationships that sustain collaborative productivity independent of brokerage function. The strongest network position combines both: bridging structural holes between dense clusters rather than maximizing either logic in isolation. This distinction is essential for interpreting the paper's central empirical finding for the time period studied. The U.S. experienced an approximately 80% decline in topological betweenness centrality while retaining substantial dominance in weighted betweenness centrality, which incorporates collaboration volume. These two measures are not contradictory; they reflect the brokerage/closure distinction directly. The U.S. lost its brokerage capital as partner nations formed direct ties with one another without U.S. intermediation. What it retained was its closure capital — deep, voluminous bilateral relationships delivering genuine collaborative value independent of the brokerage function they no longer serve. Conflating these two forms of network capital, and reading topological centrality decline as evidence of comprehensive scientific decline, is an analytical error that may be at the heart of the policy response this paper examines. We will explore this further below.

In any communications network, structural holes are unstable. Nodes on either side of a hole have incentives to connect directly, bypassing the broker and any delays or associated intermediation costs imposed by the central hub. As networks mature and densify, holes close progressively. This occurs through the accumulated effect of individual connection decisions that collectively render broker positions redundant. This is an emergent property of network evolution rather than the result of any coordinated strategy (Burt, 1992). This process generates a specific and testable empirical signature, called declining topological betweenness centrality in the unweighted network. It can be interpreted as rising global efficiency within the network as knowledge travels more directly between nodes. We can measure this by examining changes to clustering coefficients, identifiable as nodes forming direct ties, which create denser local neighborhoods. All three of these patterns are measurable in the data we present across the full 2001–2024 period.

The structural holes framework explains why U.S. brokerage advantage was vulnerable to erosion as the global network matured. The Bianconi-Barabási fitness model (2001) explains why China specifically accelerated that erosion. Standard preferential attachment strongly favors an incumbent hub, where new entrants attract connections in proportion to existing connectivity, a rich-get-richer dynamic that would favor the U.S. over any newcomer. The fitness model adds a node-level attractiveness independent of current connectivity: a high-fitness newcomer can accumulate connections at a rate disproportionate to its initial position because other nodes respond to scientific quality as well as network incumbency. China's trajectory across the 2000s and 2010s—its rapid growth in citation impact, research infrastructure, and field-specific excellence—is precisely what rising fitness describes. As Chinese institutions became increasingly compelling direct partners on scientific grounds, researchers across the network formed ties with them that bypassed U.S. intermediation, progressively closing holes the U.S. had spanned. The reorganization was not directed against the U.S. or targeted by China; it was the aggregate outcome of individual attachment decisions responding rationally to China's rising scientific attractiveness (and that of others nations, too), operating according to the network's own logic rather than political direction and consistent with what our prior work and the broader literature suggest about the autonomy of network dynamics.

**Hypothesis and Expectations**

This three-part account — preferential attachment generating the initial hub architecture, structural holes theory explaining what that position conferred and how it shifted, and the fitness model explaining China's catalytic role — generates five expectations testable using network data. 1. U.S. topological betweenness centrality will erode while weighted centrality is maintained, reflecting the brokerage/closure distinction. 2. The decline will manifest alternative pathways and new connections (rather than create a new hub), consistent with distributed hole closure. 3. Global efficiency and clustering coefficients will rise as redundant pathways multiply and denser links proliferate. 4. The reorganization will be visible consistently across scientific fields, since the change is systemic. 5. China's early network participation will predict subsequent structural change across multiple metrics and fields, consistent with its role as a high-fitness entrant whose rising scientific attractiveness catalyzed the direct connections that closed previously U.S.-spanned holes. The analyses presented below address each expectation in sequence.

**Methodology for Data Collection and Network Analysis**

This study draws on the OpenAlex database (Priem, J., Piwowar, H., & Orr, R., 2022) an open-access index of scholarly output. From this database we extracted citable publications (articles, notes, and letters) published between 2001 and 2024, yielding a corpus of 136.6 million documents, which are mainly scholarly (rather than corporate) research. Conference papers, books, and other document types were excluded to ensure comparability of collaboration indicators across the full period. Country affiliations were assigned based on the institutional addresses recorded in each publication using full counting, meaning that each country-pair collaboration is counted once per publication regardless of the number of authors from each country. Countries publishing fewer than ten internationally co-authored publications in a given year were excluded from that year's network to reduce noise from marginal participants.

Annual collaboration networks were constructed across the 24-year period, with nodes representing countries, and edges representing co-authorship links. Edge weights reflect the total number of co-authored publications between each country pair in a given year. To reduce year-to-year fluctuation and emphasize longer-term structural trends, a three-year rolling window was applied in the calculation of all network metrics reported in the analysis and figures. OpenAlex organizes its holdings in a hierarchical classification comprising four high-level domains, 36 fields, 252 subfields, and over 4,500 specific topics, assigned algorithmically based on title, abstract, and citation data, which we adopted. Our analysis was conducted at the field level, with separate annual collaboration networks constructed for all fields combined and for each of the 36 individual fields, enabling both global and field-specific analysis. All network construction, analysis, and visualization was performed using the NetworkX package in Python, with the Fruchterman-Reingold algorithm applied for network layout in visualizations. Data are available on figshare at 10.6084/m9.figshare.31567417.

The paper analyzes structural changes in the global network over 2001–2024 through betweenness centrality, k-core decomposition, community structure, and Granger causality analysis, and discusses the policy implications of these findings in an era of rising scientific nationalism.

The primary structural measure examined is betweenness centrality, which quantifies the extent to which a node lies on the shortest paths between other nodes in the network. This identifies which countries' practitioners serve as key bridges in the flow of scientific knowledge. We compute standard, normalized, and weighted variants of betweenness centrality to ensure robustness across differences in national publication volume. Additional network measures used include eigenvector centrality, k-core decomposition, clustering coefficients, community structure, and global efficiency. Full mathematical specifications for each measure are provided in the Technical Appendix.

To test the hypothesis about China's fitness, we employ Granger causality analysis examining whether China's participation in the global network, measured as the proportion of its internationally co-authored publications relative to global output, in earlier years predicted subsequent changes in global network structure. Unlike the descriptive network metrics, Granger causality analysis is conducted on raw annual time series data without rolling window smoothing, as smoothing would distort the temporal relationships the test is designed to detect. A Vector Autoregression model was estimated for each network metric across all fields combined and for individual fields, with lag lengths ranging from one to six years and optimal lag selection based on the Akaike Information Criterion. Results significant at $p \leq 0.05$ are interpreted as evidence that China's network participation Granger-causes subsequent structural change. Full details of the Granger causality procedure are provided in the Technical Appendix.

A critical consideration in interpreting our findings is that the publication data upon which our network analysis is based represents lagging indicators of scientific activity. Publications typically reflect research activities that began years earlier—collaborations may have formed, funding secured, and projects initiated as much as five years before resulting publications appear in databases. This acknowledgement of the temporal lag is essential for properly contextualizing the network reorganization we observe.

## Results

We evaluate our network analysis against five expectations presented in the theoretical framework, addressing each in sequence.

Figure 1 presents normalized betweenness centrality (panel a) and normalized weighted betweenness centrality (panel b) for the United States (U.S.), Great Britain (GB), and China from 2003 to 2024, and together the two panels reveal a picture of network restructuring. Panel (a) shows the United States in a steep and consistent decline in topological bridging roles across the entire period, falling from approximately 0.17 in 2003 to 0.035 by 2024 — a reduction of roughly 80%. Great Britain followed a similar trajectory, declining from approximately 0.085 to around 0.02. China by contrast remained essentially flat throughout, its normalized betweenness centrality barely rising from its near-zero starting position. Notably, the three nations show a pattern of downward convergence over time—not because China has risen to assume a comparable bridging role, but because the U.S. and Great Britain have declined toward the low level at which China has remained throughout. This finding is significant: the collapse of U.S. topological centrality cannot be attributed to Chinese ascendance, since China's bridging role in the unweighted network shows negligible growth. The U.S. decline is better understood as a consequence of network densification: as more countries developed direct collaborative relationships with one another, the need for a central single hub diminished systematically. The scale of this transformation is illustrated by a single statistic: in 1980, 17 nations published more than 5,000 scholarly articles in the Web of Science; by 2020 that number had grown to more than 70 nations. The emergence of this multipolar scientific landscape created the structural conditions under which hub-dependence around the U.S. became progressively less necessary or efficient.

Panel (b) shows an important qualification to the interpretation. When collaboration volume is incorporated through edge weighting, the picture shifts. The U.S. retains a dominant position throughout the period, with normalized weighted betweenness centrality remaining above 0.90 for most of the period before modest decline toward 0.91 by 2024. China however shows meaningful growth in this measure, rising from near zero to approximately 0.10 — converging with Great Britain's position by the end of the period. This suggests the U.S. has not lost its considerable role as a partner, and neither has China assumed a topological bridging role comparable to the U.S. It shows that China's actual collaborative relationships have deepened and diversified substantially but not at the expense of U.S. links. To ensure these findings are not artifacts of any particular normalization approach, we further compared weighted betweenness centrality across Log, Salton, and Jaccard normalization variants; all three consistently show a relative decline in U.S. betweenness and a relative increase for China, confirming the robustness of the measure (see Appendix). Together the two panels support the first expectation around betweenness centrality and weighted centrality. The U.S. lost its structural position as the essential intermediary in global scientific collaboration not through direct displacement but through the proliferation of alternative pathways — a process in which China's growing collaborative reach is one contributing factor among several driving broader network reorganization.

*Figure 1 Normalized betweenness centrality (2001-2024). China's rising centrality contrasts with U.S./Great Britain declines, suggesting shifting roles in global collaboration networks. A 3-year rolling average was applied to smooth short-term fluctuations and emphasize long-term trend.*

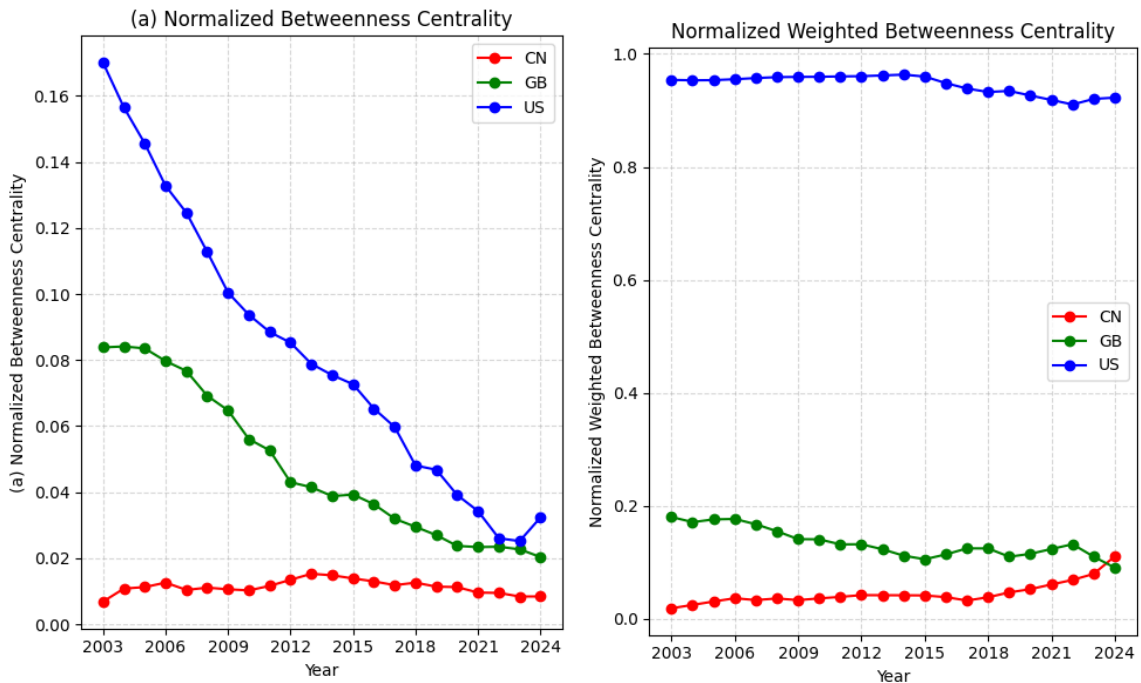

Figure 2 extends the betweenness centrality analysis through 2030 using forecast modeling, and adds two important dimensions to Figure 1. First, the historical trajectories confirm the pattern of steep and consistent U.S. and Great Britain decline alongside China's relative stability across the 2001-2024 period. A notable uptick in U.S. centrality is visible around 2023-2024, which may reflect post-pandemic recovery in international collaboration or early network responses to shifting geopolitical conditions, though its persistence remains to be seen. Second, and more consequentially, the forecast trajectories reveal strikingly different levels of uncertainty across the three nations. The U.S. forecast shows a modest potential recovery toward approximately 0.045-0.05 by 2030, but this is accompanied by an extremely wide confidence interval, suggesting that the U.S. trajectory is genuinely uncertain in ways that Great Britain and China's are not. Both Great Britain and China show narrow confidence bands and relatively flat forecasts, suggesting their network positions have stabilized at lower but predictable levels. The convergence of all three nations toward similar centrality values by 2030 is consistent with the expectation that efficiency will rise: without external shocks, the global network could evolve toward a more distributed structure in which no single nation occupies the dominant bridging position that the U.S. held. The wide uncertainty band around the U.S. forecast is itself analytically significant — it suggests that policy choices made in the near term, including decisions about international collaboration and research security, could meaningfully influence whether the U.S.stabilizes or continues to decline in network position. This has direct implications for the policy discussion we develop in a later section.

*Figure 2. Normalized Betweenness Centrality Trends and Forecasts*

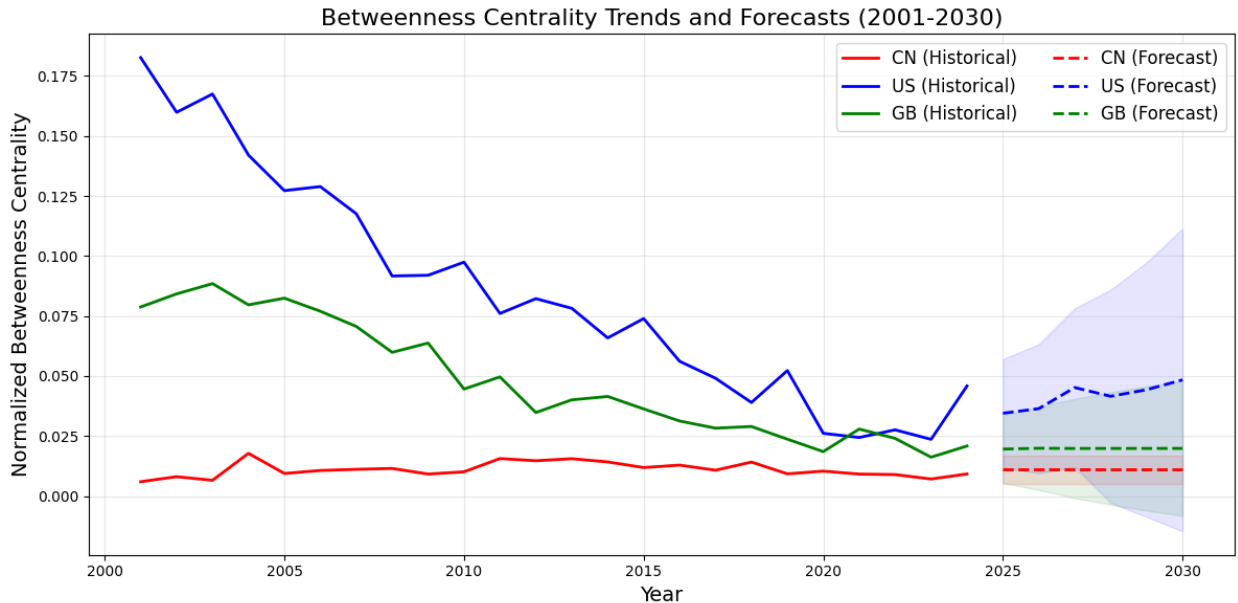

Figures 3 and 4 present visualizations of the maximum k-core collaboration network in 2001 and 2022 respectively — the most densely interconnected subnetwork of nations in which every member maintains collaborative relationships with at least k others. (The k-core reveals both how densely interconnected the collaborative center has become and how broadly participation in that center has expanded.) Node size reflects normalized betweenness centrality calculated within the k-core subnetwork, and edge thickness reflects collaboration volume. This representation is analytically distinct from the full network analysis presented in Figures 1 and 2: rather than capturing all international collaboration, it isolates the dense collaborative core where knowledge flows are most intensive and structural positions most consequential. Together the two figures provide direct visual evidence for the expectation that China's integration catalyzed systemic network reorganization through decentralized adaptation rather than direct displacement.

In 2001, the k-core network displays a pronounced hub-and-spoke architecture centered on the United States: the U.S. node dwarfs all others and its edges radiate outward to virtually every country in the core. Great Britain appears as the second largest node, with France, Germany, Canada, Australia, Japan, and Korea occupying secondary positions. China is absent from the k-core entirely at this point, reflecting its minimal presence in the densely connected core of the international collaboration network at the opening of the period.

By 2022 the structural transformation within the k-core is visually notable. The U.S. node remains the largest but Great Britain has grown substantially — a finding that reflects Great Britain's post-Brexit bridging role within the dense collaborative core, connecting the European cluster to the US-Canada-Australia axis, rather than its topological position in the full global network where its betweenness centrality has declined considerably (Figure 1). The European cluster has densified dramatically — France, Germany, Italy, Spain, Netherlands, Belgium, Sweden, Norway, and others now show strong direct interconnections among themselves within the core rather than routing primarily through the U.S. China has entered the k-core and appears as a visible node positioned between the U.S. and Great Britain, and new entrants (including India, Turkey, South Africa, Malaysia, and Poland) have joined the core, confirming the expansion of the densely connected collaborative center documented in Figure 5. The overall architecture of the k-core has shifted from hub-dependent to increasingly distributed, as expected. Critically, the U.S. has not been displaced: it remains the largest and most central node within the k-core. What has changed is that it is no longer the indispensable intermediary it was in 2001, as alternative pathways and new regional clusters have proliferated throughout the global network.

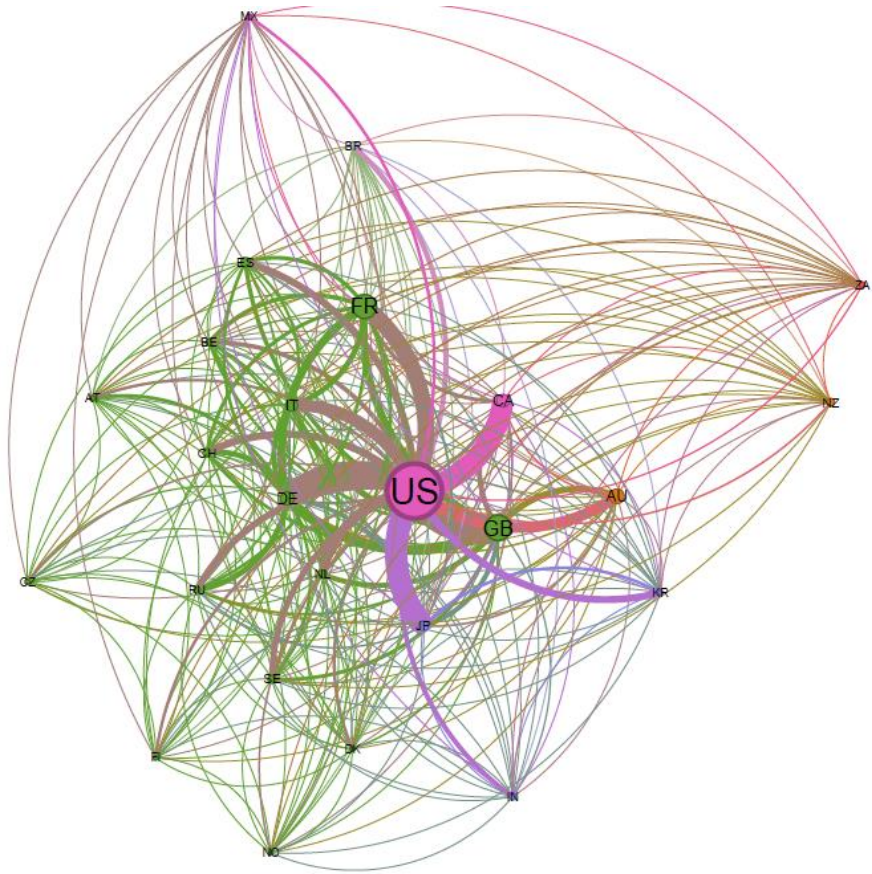

*Figure 3.. Core collaborating countries using betweenness centrality, 2001*

US
GB
FR
AU
DE
CA
IT
JP
ES
SE
CH
NL
NZ
BR
IN
BE
RU
ZA
AT
MX
KR
CZ
NO
FI

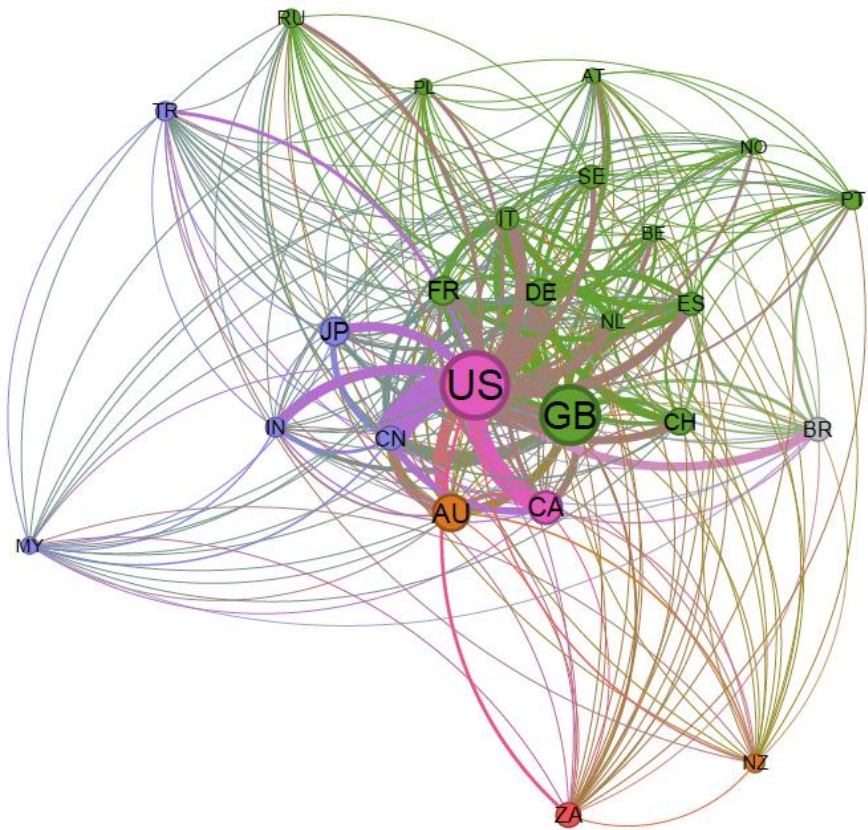

*Figure 4. Core collaborating countries using betweenness centrality measure over a three [year period] 2022*

Figure 5 provides some of the most compelling evidence in support of the expectation of the development of alternative pathways shown by examining the proportion of shortest paths between key international players that pass through either the United States or China. Panel (a) reveals that in 2003 the United States appeared in approximately 99% of the shortest paths connecting China to its international collaborators, and in approximately 89-95% of paths connecting Great Britain and Germany to theirs — confirming the U.S. as the near-total intermediary in global scientific collaboration at the opening of the period. The subsequent decline in this proportion tells a story of progressive network reorganization: European nations, particularly Britain and Germany, began building alternative pathways earlier than others—around 2013-2015 — likely reflecting the densification of intra-European collaboration networks. China's reliance onU.S.intermediation declined more sharply and later, beginning around 2017-2018, a timing that coincides with the onset of U.S.-China research security tensions, and suggests that geopolitical pressures accelerated what network dynamics had already begun. By 2024 the U.S. remained present in approximately 87% of China's shortest paths, 70% of Britain's, and 78% of Germany's — still substantial but dramatically reduced from near-total dominance two decades earlier.

Panel (b) shows the reciprocal development: China's emergence as a bridge in the shortest paths of other nations' international collaborations. Through approximately 2017, China's bridging role was essentially negligible, hovering near 0.02 for all three source nations. From 2017 onward, China's bridging role grows sharply and accelerates, reaching approximately 0.065 for Germany by 2024, while the U.S. and Britain lines remain lower and closely overlapping. Germany's particularly rapid growth in reliance on Chinese intermediation suggests that China has assumed an increasingly important role in connecting European and Asian scientific communities. Together the two panels provide direct empirical support for the expectation that the U.S.'s role as indispensable intermediary has diminished significantly as China and other nations have developed alternative bridging links, creating the expected pathways reducing hub-dependence in the network. The timing of the most significant changes—concentrated around 2017-2018—further suggests that the network reorganization was already well underway before policy interventions accelerated the changes, a point we return to in the discussion section.

*Figure 5. The U.S. or China as a bridge in the shortest path of key players, 2001-2024. A 3-year rolling window was applied.*

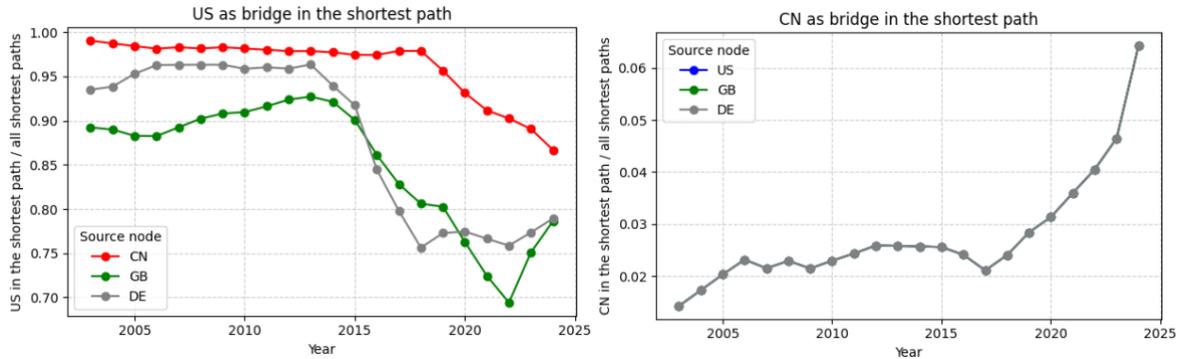

Figure 6 presents two complementary measures of network integration for China, Great Britain, and the United States across the 2003-2024 period. Panel (a) shows normalized degree centrality — the proportion of all possible collaborative connections that each nation has established — and provides the clearest visual evidence of China's expanding integration into the global scientific network. China begins the period at approximately 0.51, connecting with roughly half of all possible partner nations, and rises steeply and consistently to approximately 0.91 by 2024 — a trajectory that dwarfs the more modest gains of Great Britain and the U.S., both of which begin near saturation and converge toward near-total connectivity by the end of the period. This pattern directly supports the third prediction of hypothesis: China's entry into the network created new connection opportunities that it progressively seized, dramatically expanding its collaborative reach while the established powers approached the upper limit of possible connections. The convergence of all three nations toward near-saturation by 2024 further confirms that the global network has become substantially more inclusive and densely connected than it was at the opening of the period.

Panel (b) presents the modularity scores of each nation's ego-network — a measure of how distinctly structured the collaborative communities surrounding each nation have become. All three nations show rising modularity from 2003 through approximately 2018-2019, indicating that collaborative communities were becoming increasingly well-defined and internally cohesive throughout the period. The sharp and fully synchronized peak across all three nations around 2018-2019, followed by an equally sharp and synchronized decline, is among the most striking findings in our data. The simultaneity of this pattern across three nations with very different scientific systems and geopolitical positions strongly suggests a common external cause rather than idiosyncratic national dynamics. Given that publication data typically lags actual collaborative activity by two to three years, the behavioral changes reflected in this peak most plausibly originate around 2015-2016 — a period we examine in detail in the discussion section. A modest recovery in modularity is visible around 2022-2024 for theU.S.and GREAT BRITAIN, while China's recovery appears slightly delayed and less pronounced. The implications of this pattern are developed fully in the discussion section.

*Figure 6 The expansion of China's ego-network, 2001-2024. (a) Normalized Degree centrality, China, Great Britain, USA. (a) Modularity score of ego networks of the three nations. A 3-year rolling window was applied.*

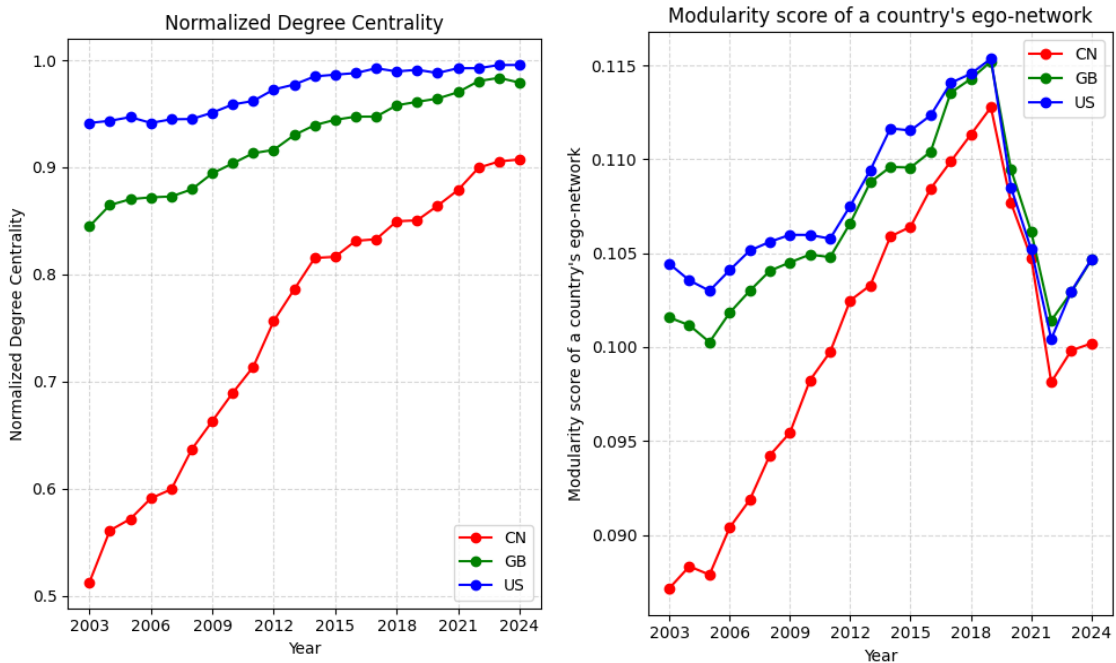

Figure 7 presents four measures of global network cohesion and efficiency that together provide robust structural evidence for the network densification predicted by the efficiency prediction informed by the hypothesis. Panel (a) shows two complementary k-core measures — the maximum k-core value and the ratio of k-core size to total network size — both of which rise consistently and almost linearly across the full 2003-2024 period. The maximum k-core more than doubles from approximately 45 to approximately 102, indicating that the most densely interconnected core of the global network has grown dramatically in absolute terms. Simultaneously the proportion of all nations participating in this dense core rises from approximately 28% to 55%, meaning that by 2024 more than half of all nations in the network are members of the densely connected core — compared to less than a third two decades earlier. This expansion of the core directly supports the third prediction of the hypothesis: peripheral nations have gained new connection opportunities and progressively joined the densely connected center of the global network, and global efficiency has risen.

Panel (b) further supports the improvement to global efficiency where we see average clustering coefficient and global efficiency, both of which rise consistently from 2003 through 2024. The clustering coefficient increases from approximately 0.77 to 0.83, indicating that nations' collaborative neighborhoods have become increasingly interconnected — countries sharing common partners are now more likely to collaborate directly with one another, forming denser local clusters. Global efficiency rises from approximately 0.60 to 0.77, indicating that knowledge can travel more directly between any two nations in the network with fewer intermediary steps— a direct structural measure of reduced hub-dependence. Both measures show notable acceleration around 2018-2021, suggesting that network densification continued and even intensified precisely during the period of greatest geopolitical tension. Taken together the two panels present a striking finding: while Figure 6 showed that ego-network modularity was significantly disrupted by geopolitical events around 2018-2019, the broader structural measures of network cohesion and efficiency continued their upward trajectory undisturbed. This divergence suggests that policy-driven disruptions affected the character of collaborative communities without reversing the deeper structural transformation of the global network toward greater density, efficiency, and distributed connectivity—a transformation driven by adaptive dynamics that operate at a more fundamental level than any single policy intervention could predict.

*Figure 7. Network cohesion and relative k-core; global efficiency,U.S.and China, 2001-2024. A 3-year rolling window was applied.*

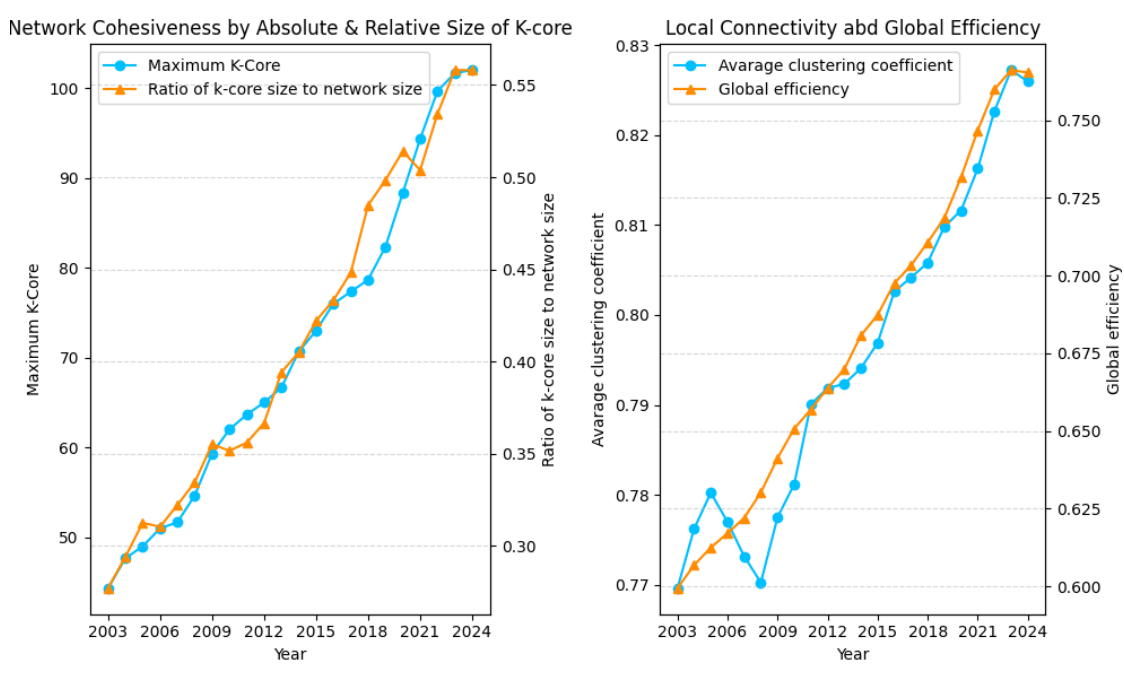

## Granger Causality Analysis: Empirical Support for China's Early Impact

Figure 8 presents results from Granger causality tests examining whether China's participation in the global network — measured as the percentage of its internationally co-authored publications — in earlier years predicted subsequent changes in global network structure. For each field, panel data covering the years 2001-2024 was used to estimate Vector Autoregression models with lag lengths of one to six years, with the minimum p-value across lag specifications reported. The heatmap displays significant p-values (≤ 0.05) highlighted in light red, with rows representing fields — row 0 denoting all fields combined and rows 11-36 representing specific scientific fields — and columns representing network metrics including clustering coefficient, k-core measures, community structure, modularity score, betweenness centralization, average betweenness centrality, and global efficiency. A cell highlighted in red indicates that in the corresponding field, China's participation in the global network predicts subsequent changes in the network metric shown in that column.

The Granger causality results are interpreted against the structural framework's fifth expectation: that China's early participation, as a high-fitness entrant, should predict subsequent structural reorganization consistently across fields. A critical interpretive constraint applies from the outset: because similar predictive relationships are observed for other major scientific nations, the analysis cannot establish that China uniquely caused the reorganization. What the results can establish is whether China's integration is consistent with the predicted systemic signature — widespread, field-crossing, and concentrated in the structural metrics most relevant to hole closure — which is what the framework requires. Several patterns in the heatmap are consistent with this expectation, with important variations across network metrics and fields.

The most striking and consistent finding concerns three network metrics—clustering coefficient, k-core nodes, and global efficiency—show significant Granger causality relationships with China's network participation across nearly every field examined. This near-universal pattern indicates that China's early integration into the global network systematically predicted subsequent improvements in local connectivity, expansion of the densely connected network core, and increases in the efficiency with which knowledge could travel between nations, consonant with the structural changes shown in Figures 6 and 7. These relationships hold consistently across fields as diverse as Agricultural and Biological Sciences, Computer Science, Mathematics, and Social Sciences, suggesting that China's integration catalyzed a broad and systemic network transformation. This finding constitutes the strongest empirical support in the analysis for the structural holes hypothesis.

The betweenness centralization column reveals a heterogeneous pattern. Significant predictive relationships between China's early participation and subsequent changes in betweenness centralization (the metric most directly relevant to the redistribution of bridging roles documented in Figures 1 and 2) appear in a subset of fields including Agricultural and Biological Sciences, Business Management and Accounting, and fields spanning Chemistry through Materials Science. This selectivity is informative: the fields where China's participation most strongly predicts subsequent shifts in the distribution of bridging roles are those where China developed substantial research capacity earliest, consistent with the hypothesis's expectation that network reorganization cascades from fields where new entrants gain capability first.

The modularity score column shows the weakest and least consistent Granger causality relationships across fields. This heterogeneous pattern is consistent with the network as a complex adaptive system in which China's integration influenced different structural dimensions through varied causal pathways and at different rates.

The significant causal relationships do not appear uniformly across all fields but rather show variation that reflects the uneven pace of China's scientific capacity development across domains. This complex pattern supports the hypothesis's conception of network reorganization as an emergent process in which cascading adaptations percolate through the network over time, with lag periods reflecting the time required for individual behavioral changes to aggregate into measurable structural shifts.

Taken together, the Granger causality results provide suggestive but limited and lag-sensitive evidence consistent with the hypothesis, indicating that China's changing network participation preceded and was temporally associated with subsequent reorganization of global network structure across multiple dimensions and fields.

*Figure 8 Granger Causality Tests Results for 2001-2024 Network Measures. For each field–metric pair, p-values are first adjusted using the Benjamini–Hochberg false discovery rate (FDR)*

*correction across all tests, and the minimum FDR-adjusted p-value across six lag specifications is reported.*

| | clustering_coefficient | k_core_nodes | modularity_score | betweenness_centralization | average_bc | global_efficiency |
|---|---|---|---|---|---|---|
| All fields | 0.0 | 0.0 | 0.0 | 0.0 | 0.002 | 0.0 |
| Agricultural and Biological Sciences | 0.054 | 0.001 | 0.002 | 0.0 | 0.0 | 0.0 |
| Arts and Humanities | 0.028 | 0.064 | 0.009 | 0.0 | 0.0 | 0.0 |
| Biochemistry, Genetics and Molecular Biology | 0.0 | 0.0 | 0.0 | 0.077 | 0.0 | 0.0 |
| Business, Management and Accounting | 0.0 | 0.0 | 0.0 | 0.0 | 0.0 | 0.0 |
| Chemical Engineering | 0.0 | 0.0 | 0.004 | 0.0 | 0.0 | 0.0 |
| Chemistry | 0.0 | 0.011 | 0.0 | 0.565 | 0.074 | 0.0 |
| Computer Science | 0.0 | 0.0 | 0.272 | 0.021 | 0.185 | 0.0 |
| Decision Sciences | 0.0 | 0.0 | 0.0 | 0.064 | 0.0 | 0.001 |
| Earth and Planetary Sciences | 0.001 | 0.001 | 0.15 | 0.03 | 0.002 | 0.002 |
| Economics, Econometrics and Finance | 0.001 | 0.0 | 0.0 | 0.0 | 0.0 | 0.101 |
| Energy | 0.0 | 0.0 | 0.367 | 0.098 | 0.0 | 0.0 |
| Engineering | 0.0 | 0.001 | 0.0 | 0.0 | 0.0 | 0.0 |
| Environmental Science | 0.0 | 0.096 | 0.003 | 0.0 | 0.0 | 0.0 |
| Immunology and Microbiology | 0.0 | 0.0 | 0.0 | 0.118 | 0.0 | 0.223 |
| Materials Science | 0.0 | 0.117 | 0.0 | 0.0 | 0.0 | 0.0 |
| Mathematics | 0.0 | 0.165 | 0.0 | 0.416 | 0.0 | 0.001 |
| Medicine | 0.0 | 0.0 | 0.0 | 0.031 | 0.161 | 0.0 |
| Neuroscience | 0.007 | 0.0 | 0.669 | 0.0 | 0.0 | 0.375 |
| Nursing | 0.0 | 0.0 | 0.0 | 0.0 | 0.485 | 0.0 |
| Pharmacology, Toxicology and Pharmaceutics | 0.027 | 0.0 | 0.0 | 0.0 | 0.0 | 0.0 |
| Physics and Astronomy | 0.135 | 0.013 | 0.0 | 0.0 | 0.0 | 0.563 |
| Psychology | 0.0 | 0.0 | 0.509 | 0.043 | 0.0 | 0.0 |
| Social Sciences | 0.0 | 0.002 | 0.0 | 0.0 | 0.0 | 0.0 |
| Veterinary | 0.051 | 0.012 | 0.0 | 0.0 | 0.005 | 0.031 |
| Dentistry | 0.209 | 0.0 | 0.002 | 0.715 | 0.0 | 0.001 |
| Health Professions | 0.0 | 0.0 | 0.578 | 0.083 | 0.0 | 0.001 |

Min FDR-adjusted p-values (≤ 0.05) highlighted in light red (IV: perc_of_intl_pubs)

Figure 9 presents normalized betweenness centrality trends across six major scientific fields — Biochemistry Genetics and Molecular Biology; Computer Science; Economics Econometrics and Finance; Environmental Science; Mathematics; and Physics and Astronomy, from 2003 to 2024. The most significant finding across all six panels is the consistency of U.S. centrality decline: without exception, the United States experienced declining bridging centrality in every field examined, confirming that network reorganization is systemic rather than concentrated in particular domains. The rate of decline varies across fields, with Computer Science, and Economics being losing share more rapidly that in Mathematics or Physics and Astronomy.

China's field-specific patterns tell a more heterogeneous story. In Environmental Science, and in Biochemistry, China's centrality remains negligible throughout, while in Mathematics and particularly Physics and Astronomy, China shows the most sustained and consistent growth, nearly converging with Great Britain by 2024 in the latter field. Great Britain's decline mirrors the U.S. pattern across all six fields, further reinforcing the interpretation that hub-dependent network structures are giving way to more distributed architectures, regardless of which nations are involved.

*Figure 9 Normalized Betweenness Centrality by Field for Six Fields, 2001-2024. A 3-year rolling window was applied.*

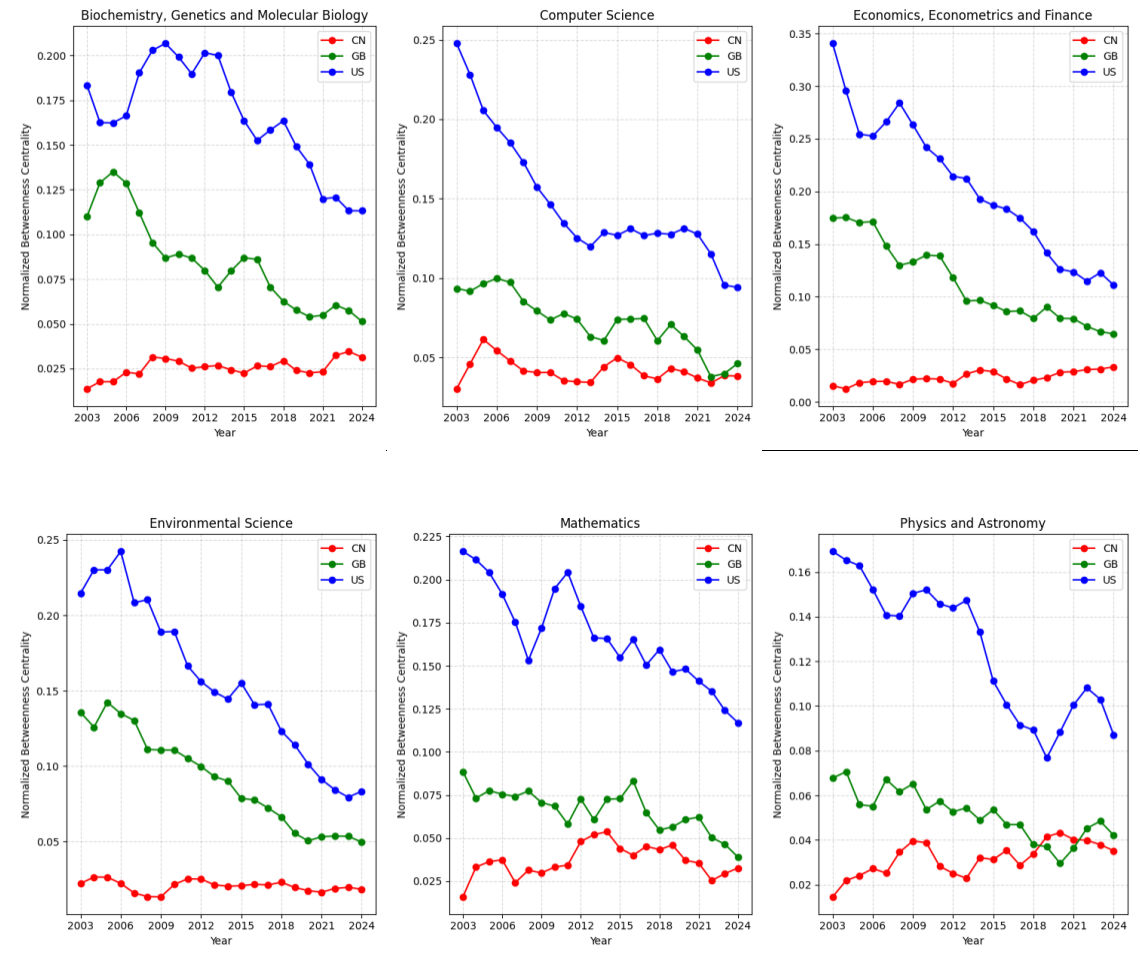

**Discussion**

The structural framework presented generated five testable expectations against which the network analysis can be evaluated. Taken together, the findings provide strong empirical support for the framework while introducing important qualifications and one significant emergent finding that the framework did not anticipate and which points toward a causal complexity the network data alone cannot resolve.

The structural expectations are strongly supported: network reorganization occurred through decentralized pathway proliferation rather than hub substitution. U.S. topological centrality declined approximately 80% while weighted centrality was retained. Global efficiency and k-core participation rose consistently — precisely the signatures of distributed hole closure.

The fourth prediction — that China's early network participation would predict subsequent structural change, tested through Granger causality analysis — is supported with qualification. The analysis demonstrates significant predictive relationships between China's early participation and subsequent changes in clustering coefficient, k-core measures, and global efficiency across multiple fields, consistent with the hypothesis that China's integration catalyzed adaptive responses throughout the network. The qualification is important: Granger causality establishes predictive temporal relationships rather than true causation, and the specific field-level patterns require careful interpretation. The finding is best understood as evidence that China's integration was associated with and preceded network reorganization in a manner consistent with an evolving network mechanism, rather than as proof of a direct causal chain.

The fifth expectation, that reorganization would be visible consistently across scientific fields, is fully supported by the field-specific analysis in Figure 9: U.S. betweenness centrality declined in each of the six major fields examined across the study period. The rate of decline varied, showing most pronounced effects in Computer Science and Economics, slowest in Biochemistry Genetics and Molecular Biology, with a uniform direction. This cross-field consistency reflects a systemic structural transformation of the global network rather than competitive dynamics concentrated in particular scientific domains.

**What the hypothesis did not anticipate**

Two findings merit further discussion as they represent departures from the hypothesis's original predictions. First, China's topological centrality—its normalized betweenness centrality in the unweighted network—did not rise substantially across the period. Figure 2 panel (a) shows China essentially flat throughout, its bridging role remaining negligible even as U.S. centrality collapsed. The hypothesis anticipated that new entrants would gain centrality as existing hubs lost it, but the data suggest the centrality released by U.S. decline was distributed diffusely across the network rather than shifting to China or any other single successor hub. This is actually a stronger confirmation of the network evolution interpretation than a simple US-to-China transfer would have been, but it requires the hypothesis to be stated more carefully than versions that imply Chinese ascendance as the mirror image of U.S. decline.

Second, and most significantly, the synchronized modularity peak and collapse across all three nations around 2018-2019 was not predicted by the structural framework and represents the analysis's most puzzling emergent finding. The framework anticipated gradual distributed hole closure, not a sharp synchronized disruption visible simultaneously across nations with very different scientific systems and geopolitical positions. This finding points toward an exogenous shock of sufficient scope to disrupt collaborative community structure across the entire network simultaneously. Critically, this shock could represent network changes related to geopolitical changes, geopolitical deterioration producing network disruption, or both operating in a feedback loop; our data cannot adjudicate among these possibilities. The appropriate analytical response to this uncertainty is not to assert a single explanation but to map the plausible causal space clearly and identify what further evidence would be required to distinguish between competing accounts — which we attempt in the policy discussion below, but which also require further study.

**The Global Turn Toward Scientific Nationalism**

The structural transformation documented in this paper is unfolding against a backdrop of significant and widening policy change across the scientifically advanced world (see James et al., 2025). Since approximately 2015, and accelerating sharply after 2018, governments across North America, Europe, and Asia-Pacific have introduced or significantly expanded frameworks variously described as research security, technological sovereignty, foreign interference prevention, and strategic autonomy in science and technology. While the specific instruments differ across national contexts they ranging from formal research security guidelines and collaboration screening mechanisms to domestic content requirements in technology investment and explicit policies of technological self-sufficiency. The policies share a common underlying logic: that scientific and technological interdependence among nations, and with China in particular, represents a vulnerability to be managed rather than a productive relationship to be cultivated.

This last point deserves emphasis because it sits at the heart of the policy paradox our analysis illuminates. The widespread institutional reluctance to engage Chinese scientific partners, to recruit Chinese researchers and students, to acknowledge Chinese scientific leadership in fields where it is demonstrably present, or to maintain collaborative relationships with China ignores the benefits of networked relationships. The global scientific community's own collaborative judgments, expressed through millions of individual decisions about partnership and co-authorship across two decades, continued to integrate China into the network core on the basis of genuine scientific capability and mutual benefit. Policy rhetoric and institutional practice have increasingly moved in the opposite direction, treating as a threat what the network's own reciprocal logic had recognized and accepted as a contribution.

The irony at the heart of this phenomenon is structural. The turn toward scientific nationalism is occurring even though the network data suggests that the distributed, networked scientific community is more efficient, more productive, and more inclusive than the hub-dependent structure dominated by the U.S. Global efficiency has risen consistently throughout the studied period. The k-core has expanded to encompass more than half of all nations. Knowledge flows more directly between any two points in the network than at any previous time in the modern scientific era. In other words, nations are retreating from a system that is functioning better than it ever has in the past. In doing so they risk forfeiting precisely the benefits that distributed participation in a more efficient network would deliver.

**Conclusion**

This paper set out to examine the structural reorganization of the global scientific network over the period 2001-2024 and to assess the implications of that reorganization for science policy in an era of rising scientific nationalism. The findings are clear, robust across multiple analytical measures, and, we argue, are consequential for how policymakers in scientifically advanced nations understand the competitive landscape they are navigating. The global scientific network underwent fundamental and irreversible structural transformation across the period examined. Using network analysis of 136.6 million publications across 24 years, we document a shift from a hub-dependent architecture centered on a small number of dominant scientific powers, to a distributed, multipolar structure in which knowledge flows more efficiently, collaborative communities are more densely connected, and participation in the network core has expanded to encompass more than half of all nations active in science. Every structural measure we examined (betweenness centrality, k-core expansion, global efficiency, clustering coefficients, degree centrality, and shortest path analysis) tells a consistent story of densification, redistribution, efficiency, and growing productivity.

Within this transformation, the United States experienced a precipitous decline in topological betweenness centrality—roughly 80% between 2003 and 2024—while retaining substantial collaborative volume as measured by weighted centrality. The distinction matters: the U.S. has lost its structural position as the essential intermediary in global scientific knowledge flows while remaining one of the most connected and productive nodes in the system. The U.S. lost the benefits of its central position in the network, but retains a robust relationship with many international partners.

China's integration contributed to this reorganization, and its scale and fitness plausibly makes its contribution qualitatively different from that of other new entrants. China's inclusion in the network is consistent with the fitness model's prediction that high-attractiveness entrants accelerate hole closure in ways that lower-fitness entrants do not. The Granger causality analysis shows that China's early network participation predicted subsequent structural change across multiple metrics and fields. The reorganization was influenced simultaneously by the growth of dozens of nations, the densification of regional clusters, and the progressive closure of structural holes through ordinary network maturation. China simply sped up the reorganization. The structural framework is supported: the network reorganized in the ways that hole closure and fitness-driven attachment predict, and the structural changes that resulted are better understood as emergent properties of network dynamics than as the intended outcomes of any single actor's policy.

**A misreading and its consequences**

The central interpretive contribution of this paper is the argument that scientifically advanced nations, and the United States in particular, have interpreted the structural transformation of the global scientific network through a framework that conflates two analytically distinct problems: the genuine and narrow problem of protecting sensitive military and dual-use research from exploitation, and the structural network problem of navigating a reorganizing landscape in which hub centrality has eroded through endogenous dynamics. The protective national policy instruments deployed appear to be calibrated primarily to the first problem while the structural evidence suggests the second may be the more consequential and the less tractable. We do not argue that all security concerns are unfounded, since documented cases of IP appropriation and the exploitation of university research relationships were real. We argue that the policy response was broader than those cases warranted, and that the excess restriction addressed a political mechanism when the network changes are a better explanation for a sense of vulnerability.

One interpretation is that the loss of hub centrality would be experienced by the U.S. system as a disorienting loss of access, visibility, and gravitational pull of the advantages that hub position had conferred. China, as the most visible new actor in the network during the period of steepest decline, became the proximate explanation for a loss whose actual causes were distributed across the entire network reorganization.

The costs visible in our data are real if not yet fully measurable. Collaborative communities disrupted around 2018-2019 have only partially recovered, as Figure 6 panel (b) shows. Co-authorship between formerly central scientific nations and China has declined in precisely the fields—advanced technologies, computational sciences, engineering—where collaborative productivity is most demonstrably valuable (Flynn et al., 2024). Whether nations that have maintained broader engagement with China have fared better than those that have aggressively restricted them is an empirical question this paper cannot answer with the data available. What our data establish is that the network disruption occurred, that it is ongoing, and that the structural case for broad disengagement is not supported by our analysis.

**The inverse policy imperative**

Hub centrality was self-reinforcing: being the dominant node in a global network meant the world's best ideas, most talented researchers, and most important developments flowed toward the U.S. almost automatically. That gravitational logic no longer operates in a distributed network. The frontier of global science is now genuinely multipolar, located in Chinese laboratories, European consortia, emerging Asian scientific powers, and research communities across the global south that were negligible actors two decades ago. A nation that retreats from that distributed landscape does not protect itself from competition, it blinds itself to developments occurring at nodes to which it is no longer connected.

The strategic imperative in a distributed network is therefore the inverse of the imperative that hub centrality created. Rather than attracting the world to a center, scientifically advanced nations must now go to the world, actively scanning a distributed global frontier for the most important ideas wherever they emerge, strategically linking with field leaders regardless of their national location, and maintaining the institutional conditions that ensure knowledge gained anywhere in the global network can find its way back to strengthen domestic scientific capacity. This requires not merely a different science policy but a fundamentally different strategic orientation, one that treats broad international connection as a competitive necessity rather than a security vulnerability, and one that recognizes the network's own preferential attachment logic as an instrument of national scientific strategy rather than a threat to it.

This reframing has concrete implications. Immigration and visa policies that restrict the movement of scientific talent sever precisely the connections through which knowledge flows in a distributed network. Research security frameworks applied broadly on the basis of national origin rather than specific security-relevant behavior reduce the connectivity on which scientific competitiveness depends while doing little to address the narrow category of genuinely sensitive research where restriction is warranted. Rhetoric that treats international scientific engagement as inherently risky deters the collaborative relationships through which scientifically advanced nations maintain their visibility into a distributed global frontier — relationships that, as our data show, the network's own preferential attachment logic had built on the basis of recognized scientific merit and mutual benefit.

**Future research**

The findings presented here raise a natural counterfactual question that the current data and design cannot directly answer: what would the global scientific network have looked like had China's integration not occurred, or had China followed a different developmental trajectory? Addressing this question rigorously would require either a node-removal analysis comparing network structural properties across actual and China-absent configurations, a synthetic control approach using comparable high-growth scientific nations as donors, or an agent-based simulation calibrated to pre-2001 network conditions and run with and without a high-fitness entrant of China's scale. Each approach carries distinct assumptions and limitations — particularly regarding the endogeneity of China's scientific rise to the network itself, and the impossibility of holding geopolitical context constant while removing a major network actor. What such analyses could establish, however, is whether the structural transition we document was accelerated by China's integration or would have occurred through ordinary network maturation regardless — a distinction with direct consequences for how policymakers should interpret the reorganization and calibrate their responses to it.

Acknowledgements

We thank Andrew Plume and others at Elsevier for help with data. We used Claude.ai to help with structuring the outline for the paper, checking grammar, and creating the reference list. We thank Anna Lise Ahlers, head of the Lise Meitner Research Group at the MPIWG for help in defining the scope of the study, and to participants in a Max Planck Institute for the History of Science workshop for comments and interactions on this topic.

# References

Adams, J., & Szomszor, M. (2022). A converging global research system. *Quantitative Science Studies*, *3*(3), 553–569.

Bornmann, L., Adams, J., & Leydesdorff, L. (2018). The negative effects of citing with a national preference in nanotechnology. *Scientometrics*, *116*(2), 1201–1212. https://doi.org/10.1007/s11192-018-2762-8

Börner, K., Maru, J. T., & Goldstone, R. L. (2004). The simultaneous evolution of author and paper networks. *Proceedings of the National Academy of Sciences*, *101*(Suppl. 1), 5266–5273.

Clauset, A., Newman, M. E., & Moore, C. (2004). Finding community structure in very large networks. *Physical Review E*, *70*(6), 066111. https://doi.org/10.1103/PhysRevE.70.066111

Collaco, J., St. Geme, J. S., Abman, S., & Furth, S. (2022). It takes a team to make team science a success: Career development within multicenter networks. *Jornal de Pediatria*, *98*(3), 227–229.

Docampo, D., & Bessoule, J. J. (2019). A new approach to the analysis and evaluation of the research output of countries and institutions. *Scientometrics*, *119*(2), 1207–1225.

Eberle, J., Stegmann, K., Barrat, A., Fischer, F., & Lund, K. (2021). Initiating scientific collaborations across career levels and disciplines: A network analysis on behavioral data. *International Journal of Computer-Supported Collaborative Learning*, *16*(2), 263–294.

Flynn, R., Glennon, B., Murciano-Goroff, R., & Xiao, J. (2024). *Building a wall around science: The effect of U.S.–China tensions on international scientific research* (NBER Working Paper No. 32622). National Bureau of Economic Research.

Forchino, M. V., & Torres Salinas, D. (2025). The OpenAlex database in review: Evaluating its applications, capabilities, and limitations. https://doi.org/10.5281/zenodo.17357948

Freeman, L. C. (1977). A set of measures of centrality based on betweenness. *Sociometry*, *40*(1), 35–41. https://doi.org/10.2307/3033543

Frenken, K., Hölzl, W., & Vor, F. (2005). The citation impact of research collaborations: The case of European biotechnology and applied microbiology (1988–2002). *Journal of Engineering and Technology Management*, *22*(1–2), 9–30. https://doi.org/10.1016/j.jengtecman.2004.11.002

Frenken, K., Ponds, R., & Van Oort, F. (2010). The citation impact of research collaboration in science-based industries: A spatial-institutional analysis. *Papers in Regional Science*, *89*(2), 351–371. https://doi.org/10.1111/j.1435-5957.2010.00309.x

Gui, Q., Liu, C., & Du, D. (2025). Global scientific collaboration patterns and trends: A comprehensive analysis of international co-authorship networks. *Research Policy*, *54*(2), 104–118.

Gunn, A., & Johansen, J. (2022). International research collaboration in the digital age: Challenges and opportunities. *Science and Public Policy*, *49*(3), 412–428. https://doi.org/10.1093/scipol/scac015

Howells, J. (1990). The internationalization of R & D and the development of global research networks. *Regional Studies*, *24*(6), 495–512. https://doi.org/10.1080/00343409012331346174

Hu, A. G. Z. (2020). Public funding and the ascent of Chinese science: Evidence from the National Natural Science Foundation of China. *Research Policy*, *49*(5), 103944. https://doi.org/10.1016/j.respol.2020.103944

James, A., Flanagan, K. ; Naisbitt, A., Rigby, J. (2025) European Research Security: Threat Perspectives and the Responses of Policy Makers. University of Manchester.

Jefferson, O. A., Kögler, D. F., Cavalli, E., & Aboy, M. (2024). International collaboration and knowledge production in biotechnology: A network analysis approach. *Technological Forecasting and Social Change*, *198*, 122945. https://doi.org/10.1016/j.techfore.2023.122945

Jeong, H., Néda, Z., & Barabási, A. L. (2003). Measuring preferential attachment in evolving networks. *EPL (Europhysics Letters)*, *61*(4), 567–572.

Jeong, S., Choi, J. Y., & Kim, J. Y. (2013). On the drivers of international collaboration: The impact of informal communication, familiarity, trust, and homophily. *Technological Forecasting and Social Change*, *80*(8), 1554–1563. https://doi.org/10.1016/j.techfore.2013.03.003

Jervis, R. (1978). Cooperation under the security dilemma. *World Politics*, *30*(2), 167–214.

Katz, J. S., & Hicks, D. (1997). How much is a collaboration worth? A calibrated bibliometric model. *Scientometrics*, *40*(3), 541–554. https://doi.org/10.1007/BF02459299

Kontopoulos, K. M. (1993). *The logics of social structure*. Cambridge University Press.

Kwiek, M. (2015). The internationalization of research in Europe: A quantitative study of 11 national systems from a micro-level perspective. *Journal of Studies in International Education*, *19*(4), 341–359. https://doi.org/10.1177/1028315315572928

Leydesdorff, L., & Wagner, C. S. (2008). International collaboration in science and the formation of a core group. *Journal of Informetrics*, *2*(4), 317–325. https://doi.org/10.1016/j.joi.2008.07.003

Leydesdorff, L., Wagner, C. S., Park, H. W., & Adams, J. (2013). International collaboration in science: The global map and the network. *El Profesional de la Información*, *22*(1), 55–61.

Li, J. T. (2018). On the advancement of highly cited research in China: An analysis of the Highly Cited database. *PLoS ONE*, *13*(8), e0201764.

Lin, W. C., & Chang, C. W. (2022). The influence of Chinese scholars on global research. *Scientific Reports*, *12*, 18410. https://doi.org/10.1038/s41598-022-23024-z

Merton, R. K. (1957). Priorities in scientific discovery: A chapter in the sociology of science. *American Sociological Review*, *22*(6), 635–659. https://doi.org/10.2307/2089193

Milojević, S. (2010). Modes of collaboration in modern science: Beyond power laws and preferential attachment. *Journal of the American Society for Information Science and Technology*, *61*(7), 1410–1423. https://doi.org/10.1002/asi.21331

Milojević, S. (2014a). Network analysis and indicators. In Y. Ding, R. Rousseau, & D. Wolfram (Eds.), *Measuring scholarly impact* (pp. 91–114). Springer. https://doi.org/10.1007/978-3-319-10377-8_3

Milojević, S. (2014b). Principles of scientific research team formation and evolution. *Proceedings of the National Academy of Sciences*, *111*(11), 3984–3989. https://doi.org/10.1073/pnas.1309723111

Monge, P. R., & Contractor, N. S. (2003). *Theories of communication networks*. Oxford University Press.

National Academies of Sciences, Engineering, and Medicine. (1999). *Evaluating federal research programs: Research and the Government Performance and Results Act*. National Academies Press. https://www.ncbi.nlm.nih.gov/books/NBK547315/

Ostrom, E. (2000). Collective action and the evolution of social norms. *Journal of Economic Perspectives*, *14*(3), 137–158. https://doi.org/10.1257/jep.14.3.137

Padgett, J. F., & Powell, W. W. (2012). *The emergence of organizations and markets*. Princeton University Press.

Pan, R. K., Kaski, K., & Fortunato, S. (2012). World citation and collaboration networks: Uncovering the role of geography in science. *Scientific Reports*, *2*, 902.

Pinheiro, F. L., Santos, F. C., & Pacheco, J. M. (2016). Linking individual and collective behavior in adaptive social networks. *Physical Review Letters*, *116*(12), 128702. https://doi.org/10.1103/PhysRevLett.116.128702

Priem, J., Piwowar, H., & Orr, R. (2022). OpenAlex: A fully-open index of scholarly works, authors, venues, institutions, and concepts. *arXiv*. https://arxiv.org/abs/2205.01833

Rosvall, M., & Bergstrom, C. T. (2007). Maps of random walks on complex networks reveal community structure. *Proceedings of the National Academy of Sciences*, *105*(4), 1118–1123.

Sá, C., & Sabzalieva, E. (2018). Scientific nationalism in a globalizing world. In C. Marope, P. J. Wells, & E. Hazelkorn (Eds.), *Rankings and accountability in higher education: Uses and misuses* (pp. 149–166). UNESCO Publishing.

Salter, A. J., & Martin, B. R. (2001). The economic benefits of publicly funded basic research: A critical review. *Research Policy*, *30*(3), 509–532.

Shu, F., Julien, C.-A., & Larivière, V. (2018). Does the Web of Science accurately represent Chinese scientific performance? *Journal of the Association for Information Science and Technology*, *69*(10), 1273–1280.

Simon, H. A. (1962). The architecture of complexity. *Proceedings of the American Philosophical Society*, *106*(6), 467–482.

Tang, L. (2013). Does "birds of a feather flock together" matter: Evidence from a longitudinal study on US–China scientific collaboration. *Journal of Informetrics*, *7*(3), 577–587.

Tang, L., Shapira, P., & Youtie, J. (2015). Is there a clubbing effect underlying Chinese research citation increases? *Journal of the Association for Information Science and Technology*, *66*(9), 1923–1932.

Wagner, C. S., & Leydesdorff, L. (2005). Network structure, self-organization, and the growth of international collaboration in science. *Research Policy*, *34*(10), 1608–1618. https://doi.org/10.1016/j.respol.2005.08.002

Wagner, C. S., Park, H. W., & Leydesdorff, L. (2015). The continuing growth of global cooperation networks in research: A conundrum for national governments. *PLoS ONE*, *10*(7), e0131816. https://doi.org/10.1371/journal.pone.0131816

Wagner, C. S., Brahmakulam, I., Jackson, B., Wong, A., & Yoda, T. (2018). *Science and technology collaboration: Building capacity in developing countries?* RAND Corporation.

Wagner, C. S., Whetsell, T. A., & Mukherjee, S. (2019). International research collaboration: Novelty, conventionality, and atypicality in knowledge recombination. *Research Policy*, *48*(5), 1260–1270. https://doi.org/10.1016/j.respol.2019.01.002

Wagner, C. S., Zhang, L., & Leydesdorff, L. (2022). A comparison of Chinese scientific performance in different fields using citation impact indicators. *Scientometrics*, *127*(4), 2089–2108.

Wu, L., Wang, D., & Evans, J. A. (2019). Large teams develop and small teams disrupt science and technology. *Nature*, *566*(7744), 378–382. https://doi.org/10.1038/s41586-019-0941-9

Xafis, V., Schaefer, G. O., Labude, M. K., Zhu, Y., & Hsu, L. Y. (2021). The challenge of global health research partnerships in the context of the COVID-19 pandemic. *Developing World Bioethics*, *21*(2), 74–78. https://doi.org/10.1111/dewb.12304

Yin, L., Kretschmer, H., Hanneman, R. A., & Liu, Z. (2006). Connection and stratification in research collaboration: An analysis of the COLLNET network. *Information Processing & Management*, *42*(6), 1599–1613.

Zhang, C., Bu, Y., Ding, Y., & Xu, J. (2018). Understanding scientific collaboration: Homophily, transitivity, and preferential attachment. *Journal of the Association for Information Science and Technology*, *69*(1), 72–86. https://doi.org/10.1002/asi.23916

Zhu, Y. P., & Park, H. (2022). Profiling the most highly cited scholars from China: Who they are. To what extent they are interdisciplinary. *El Profesional de la Información*, *31*(2), e310213.

# Technical Appendix

## The construction of network

We draw upon 136.6 million articles, notes, letters, and reviews published in journals to assess international coauthorship patterns, knowledge dynamics, and impact. The global collaboration networks for all fields combined and each specific field by year were constructed, with links between nodes denote the co-authorship of authors from any two countries in the same publication. Integer counting was applied that a publication with any authors from country A and B contributed to 1 collaboration between the two countries. As shown in Figure A1, the network has expanded throughout the period with more involving countries, collaborated links among them, and growing density.

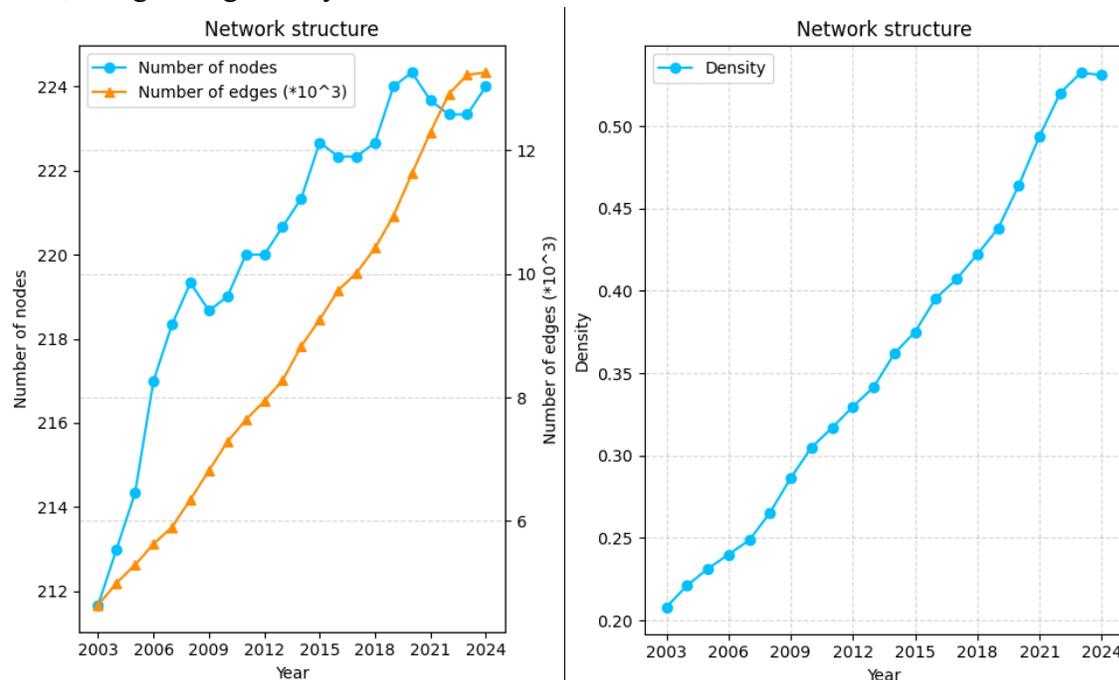

*Figure A1. Overall network metrics. A 3-year rolling average is used.*

Since edge weights are highly skewed and dependent on the productivity levels of countries, different normalization approaches – Log, Salton, and Jaccard – were applied to adjust the edge weights. This enables a more robust comparison of the weighted betweenness centrality of key actors in the network.

$$w_{log} = \log(w + 1)$$
$$w_{salton} = \frac{w}{\sqrt{p_i p_j}}$$
$$w_{jaccard} = \frac{w}{p_i + p_j - w}$$

$p_i$ and $p_j$ indicate the productivity of i and j, respectively.

## Modularity and number of communities

The number of communities is generated using NetworkX package in Python. It refers to the number of Greedy modularity communities using Clauset-Newman-Moore greedy modularity maximization approach (Clauset et al. 2004). Greedy modularity maximization begins with each node in its own community and joins the pair of communities that most increases modularity until no such pair exists or until number of communities is reached.

The modularity score, generated using NetworkX package in Python, measures the strength of division of a network into communities. It is based on community partitions generated through greedy modularity maximization and quantifies how well nodes within communities are interconnected. In other words, high modularity means more edges within communities than would be expected by chance, suggesting a strong community structure. Conversely, low modularity reflects a higher proportion of inter-community edges than intra-community edges, indicating a less cohesive network.

## Betweenness centrality

Betweenness centrality measures the extent to which a node lies on the shortest paths between other nodes in the network (Freeman, 1977). For a node $i$, the *betweenness centrality* ($BC_i$) is calculated as:

$$BC(i) = \sum_{s \neq i \neq t, \ s,t \in V} \frac{\sigma_{st}(i)}{\sigma_{st}} \quad (1)$$

where V is the set of all nodes in the network, $\sigma_{st}$ is the total number of shortest paths between nodes $s$ and $t$, and $\sigma_{st}(i)$ is the number of those paths that pass through node $i$. This measure is topology-dependent and agnostic to edge weights, as it only considers the binary existence of links between nodes, not their weights.

To compare between networks of different sizes, the *normalized betweenness centrality* is commonly used by normalizing the standard betweenness centrality by the maximum possible value in the network of $n$ nodes:

$$BC_{norm}(i) = \frac{BC(i)}{(n-1)(n-2)/2} \quad (1)$$

where $(n-1)(n-2)/2$ represents the total number of ordered node pairs excluding i, namely, the maximum number of potential paths a node could mediate in a network of $n$ nodes.

The *weighted betweenness centrality* is an extension of standard betweenness centrality that counts for edge weights in a network. It provides a more realistic measure of node importance considering the weighted shortest path between two nodes which minimizes the sum of inverse weights. For example, consider a network with edge weights $w_{AB} = 100, w_{BC} = 100, w_{AC} = 1$, the corresponding distances are: $d_{AB} = d_{BC} = 1/100, d_{AC} = 1$. The shortest path from A to C is A-B-C (d = 0.02), outperforming the direct path A-C (d=1). The *weighted betweenness centrality* of B thus accounts for such mediated paths. *Normalized Weighted Betweenness Centrality* follows the same normalization procedure as the unweighted variant, ensuring scale invariance.

Generally, nodes with high betweenness centrality are often considered important gatekeepers or bridges, as they facilitate the flow of information, resources, or influence between different parts of the network. Here, we will examine betweenness centrality scores across the years to test if China is become more central to the overall network and by field. Central nodes are likely to be critical in the overall network structure, as they connect different clusters or communities within the network. Changes in their betweenness centrality over time may indicate shifts in the network's structure or the emergence of new influential nodes.

### The shortest path

In a weighted network, the shortest path between two nodes is defined as the path with the minimum total distance, where distance is typically the inverse of edge weight. For example, consider a network with edge weights $w_{AB} = 100, w_{BC} = 100, w_{AC} = 1, w_{AD} = 2$, the corresponding distances are: $d_{AB} = d_{BC} = \frac{1}{100}, d_{AC} = 1, d_{AD} = 1/2$. The shortest paths from node D to nodes A, B, C are D-A, D-A-B, and D-A-B-C, with the total distance of 0.5, 0.51, 0.52, respectively. This illustrates that node A is crucial to node D, as it lies on all shortest paths from D to the other nodes in the network.

### Eigenvector centrality

Eigenvector centrality measures the influence of a node in a network by considering not only its own connections but also the importance of the nodes it connected to. The centrality score of a node is proportional to the sum of the centrality scores of all its neighbors. This can be mathematically represented as:

$$x_i = \frac{1}{\lambda}\sum_i^n A_{ij}x_j, \text{ or in the matrix form } A\boldsymbol{x} = \lambda\boldsymbol{x}$$

Where $x_i$ denotes the centrality of node i, $A_{ij}$ denotes the adjacency matrix of the graph G; $\lambda$ is the largest eigenvalue of A, and $\boldsymbol{x}$ is the corresponding right eigenvector. In essence, eigenvector centrality measures the extent to which a node is connected to other highly central nodes, thus reflecting both direct and indirect influence within the network.

### Granger causality

Granger causality is a statistical concept used to determine whether one time series can be useful in forecasting another. It's important to note that Granger causality doesn't imply true causation, but rather a predictive temporal relationship. Given the trending nature of several variables, all series are first-differenced to reduce potential non-stationarity, although some residual non-stationarity may remain due to the limited time span of the data (23.8% to 43.9%).

In the context of analyzing a country's influence on the global scientific network, here's how we might apply Granger causality tests. We begin by constructing time-series data that include China's internationally collaborated productivity as a percentage of global output, along with a set of global network metrics – such as *the average clustering coefficient, number of countries in the k-core, the ratio of k-core size to network size, number of communities, modularity score, betweenness centralization, average betweenness centrality, and global efficiency.* These metrics are compiled for both all fields combined and for individual field. For each dataset, we test whether changes in China's scientific output Granger-causes changes in global scientific network characteristics. Models are estimated using lag lengths ranging from 1 to 6, with the optimal lag selected based on the Akaike Information Criteria (AIC). However, given the relatively short time series (2001–2024), higher-order lag specifications substantially reduce the effective sample size, and results may therefore be sensitive to lag selection.

We would typically use a Vector Autoregression (VAR) model, which might look like this:

$$Y_t = \alpha + \beta_1 Y_{t-1} + \cdots + \beta_p Y_{t-p} + \gamma_1 X_{t-1} + \cdots \gamma_p X_{p-1} + \varepsilon_t$$

Where $Y_t$ is the global metric at time t; $X_t$ is the country's metric at time t; and p is the number of lags included.

To account for multiple comparisons across fields, metrics, and lag specifications, p-values are adjusted using the **Benjamini–Hochberg false discovery rate (FDR) correction**. For each field–metric pair, the minimum FDR-adjusted p-value across the six lag specifications is reported, representing the strongest observed lag-specific association. The p-values of the coefficients in the best-fit models are then examined, with a value below 0.05 indicating statistically significance. Given the combination of multiple testing, lag selection, and short time series, these results should be interpreted as exploratory and indicative of potential temporal associations rather than robust predictive relationships. If the adjusted p<0.05, we interpret this as evidence that the country's scientific output precedes and is temporally associated with subsequent changes in global scientific trends.

By applying Granger causality tests, we can gain insights into whether changes in a country's scientific output tend to precede changes in global scientific trends. This could provide evidence of the country's influence or leadership in global science.

Clauset, A., Newman, M. E., & Moore, C. (2004). Finding community structure in very large networks. *Physical Review E, 70*(6), 066111. https://doi.org/10.1103/PhysRevE.70.066111

# Appendix

*Figure A2. Normalized Between Centrality using different normalization approach, 2001-2024. (a) Raw values: edge weights represent the unadjusted collaboration frequencies between countries. (b) Log-transformed: edge weights are log-transformed (after adding 1 to avoid zero values). (c) Salton's cosine measure (association strength): edge weights are normalized by dividing by the square root of product of the two nodes' collaboration volumes. (d) Jaccard index: edge weights are normalized by dividing by their non-overlapping total of the two nodes' collaborations. A 3-year moving window was applied. Across four approaches, the results consistently show a relative decline in the US and a rise in China, despite differences in scale and trajectory.*

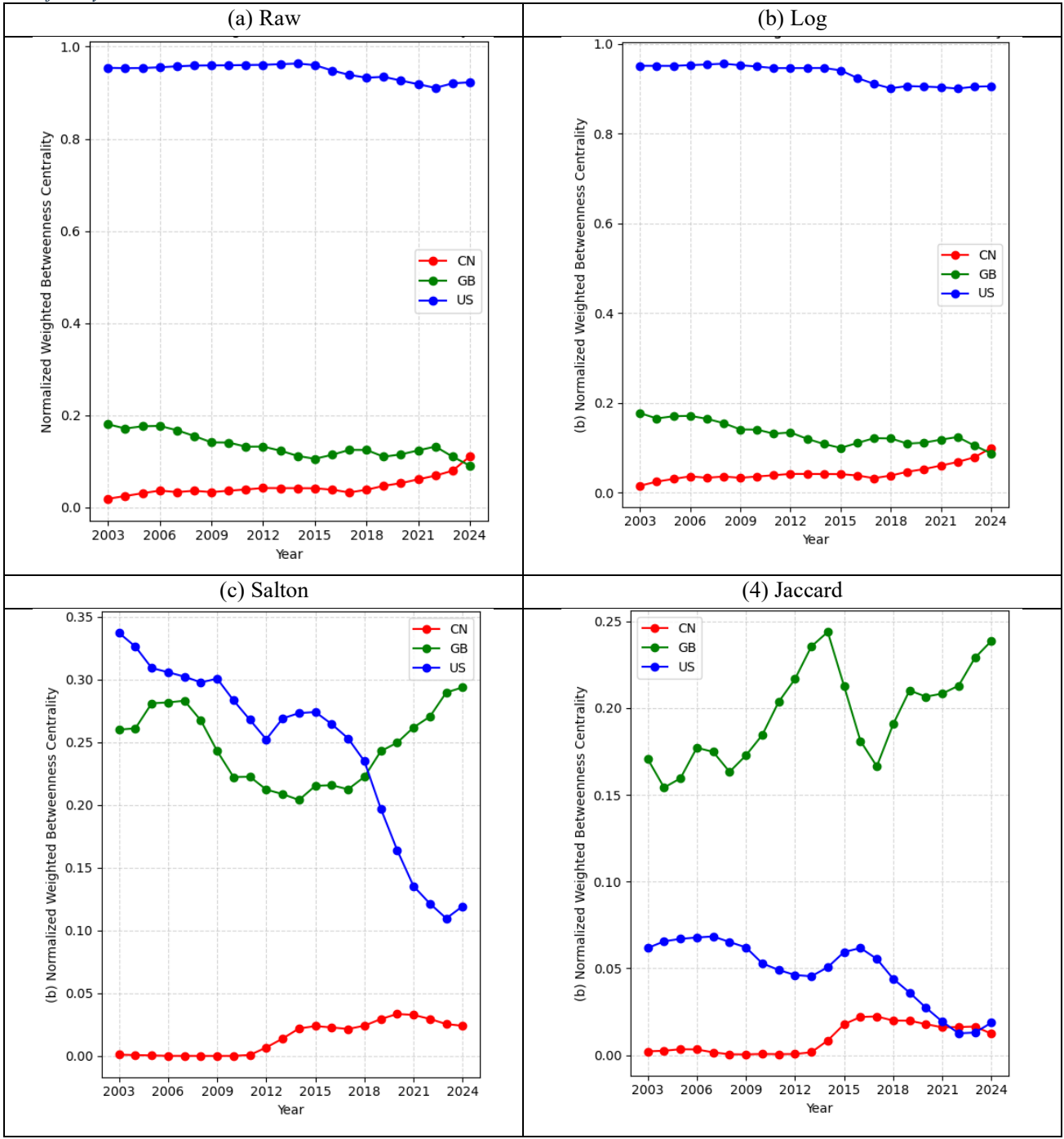

*Figure A3. Eigenvector Centrality of China, the UK, and the US, 2001-2024. A 3-year moving window was applied. The trends of the three nations indicate a steadily rising role for China in the global network, despite a slight decline in recent years. By contrast, both the US and UK exhibit a consistent downward trend, particularly pronounced for the US. These contrasting patterns point to a convergence, highlighting the influence of China's growing participation in the network on other nations.*

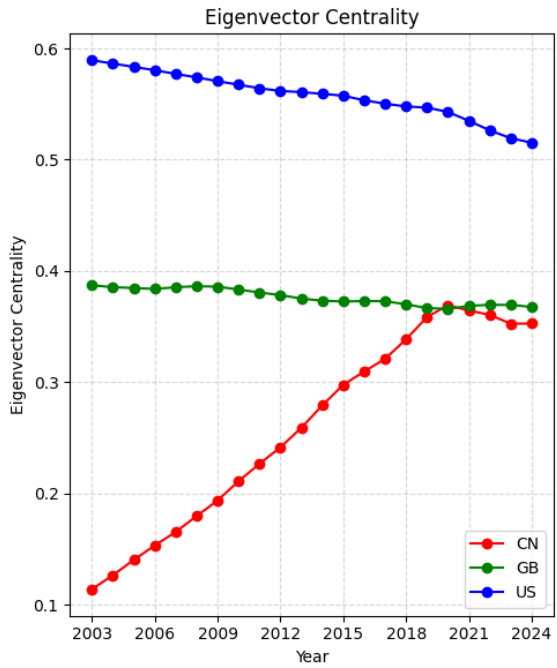

*Figure A4. Normalized Weighted Between Centrality for selected fields, 2001-2024. A 3-year moving window was applied.*

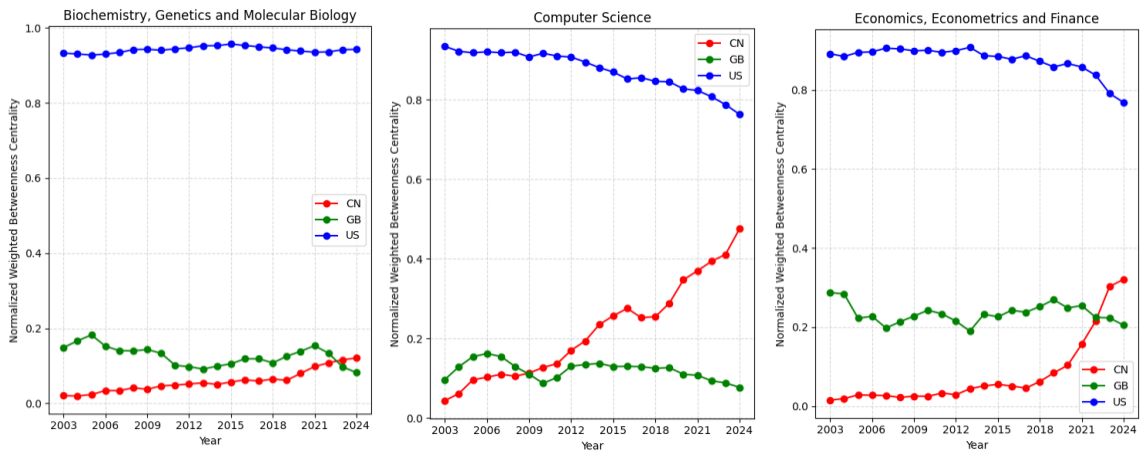

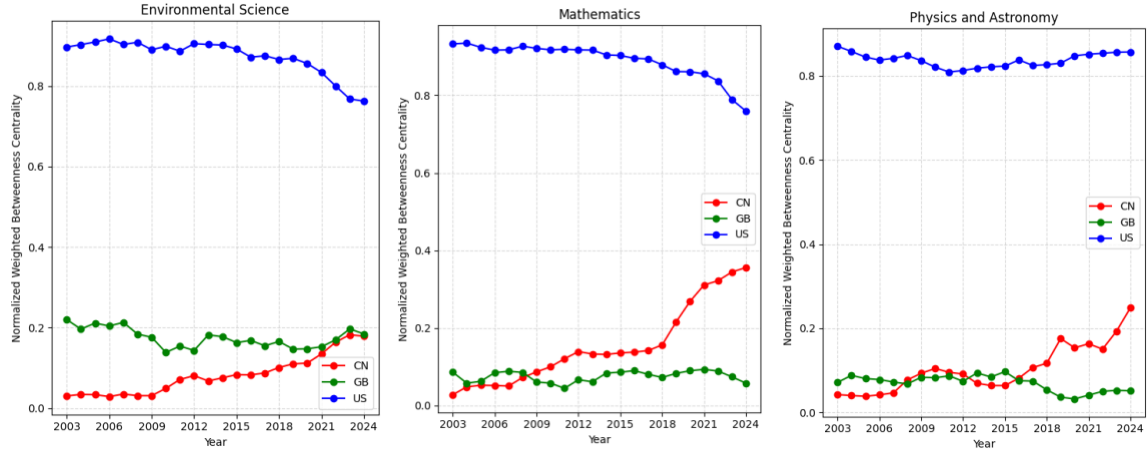

*Figure A2. Normalized Between Centrality using different normalization approach, 2001-2024. (a) Raw values: edge weights represent the unadjusted collaboration frequencies between countries. (b) Log-transformed: edge weights are log-transformed (after adding 1 to avoid zero values). (c) Salton's cosine measure (association strength): edge weights are normalized by dividing by the square root of product of the two nodes' collaboration volumes. (d) Jaccard index: edge weights are normalized by dividing by their non-overlapping total of the two nodes' collaborations. A 3-year moving window was applied. Across four approaches, the results consistently show a relative decline in the U.S. and a rise in China, despite differences in scale and trajectory.*

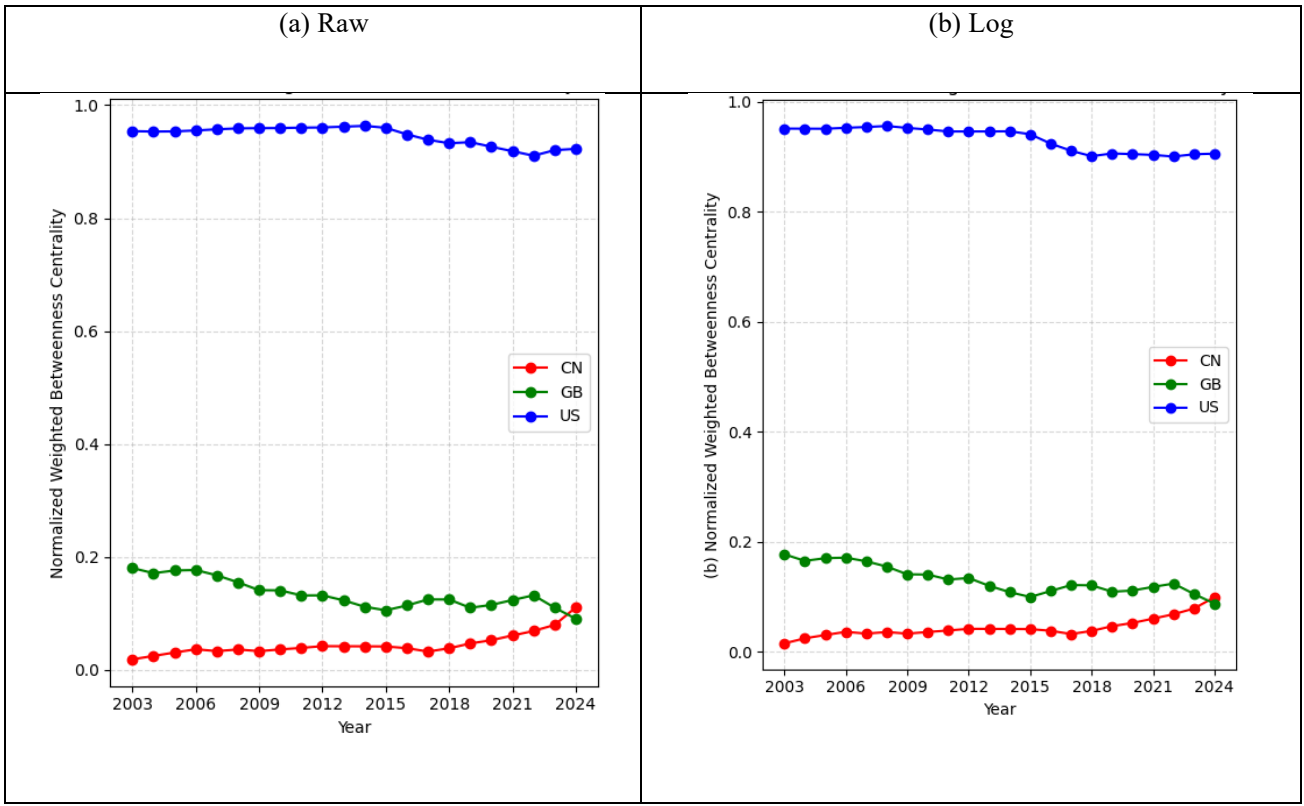

| (c) Salton | (4) Jaccard |
|---|---|

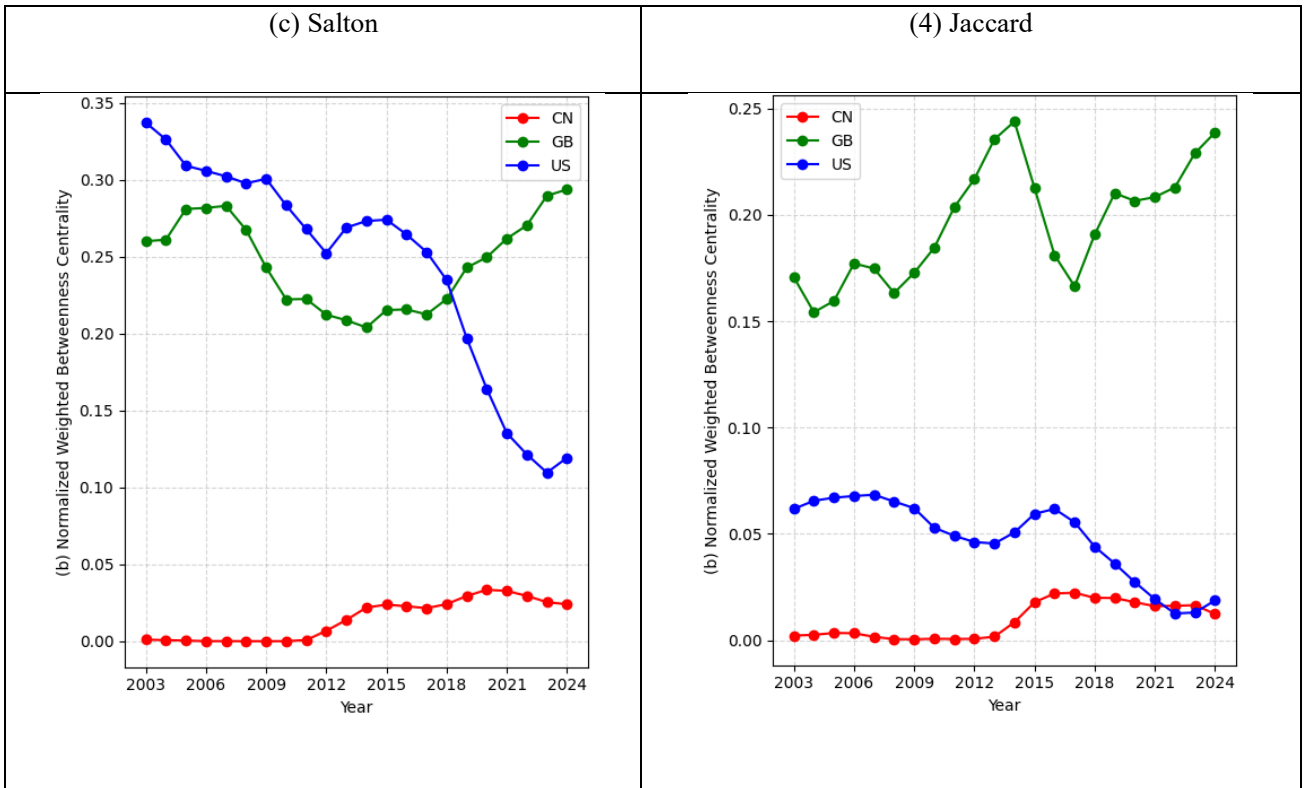

*Figure A3. Eigenvector Centrality of China, the UK, and the US, 2001-2024. A 3-year moving window was applied. The trends of the three nations indicate a steadily rising role for China in the global network, despite a slight decline in recent years. By contrast, both the U.S. and UK exhibit a consistent downward trend, particularly pronounced for the US. These contrasting patterns point to a convergence, highlighting the influence of China's growing participation in the network on other nations.*

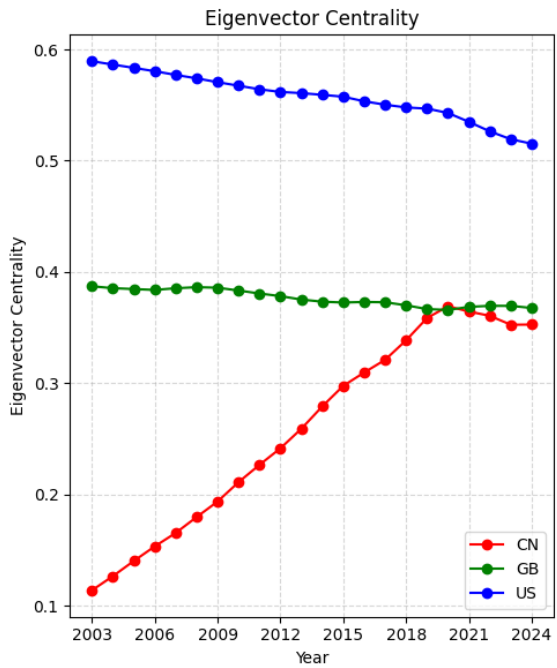

*Figure A4. Normalized Weighted Between Centrality for selected fields, 2001-2024. A 3-year moving window was applied.*

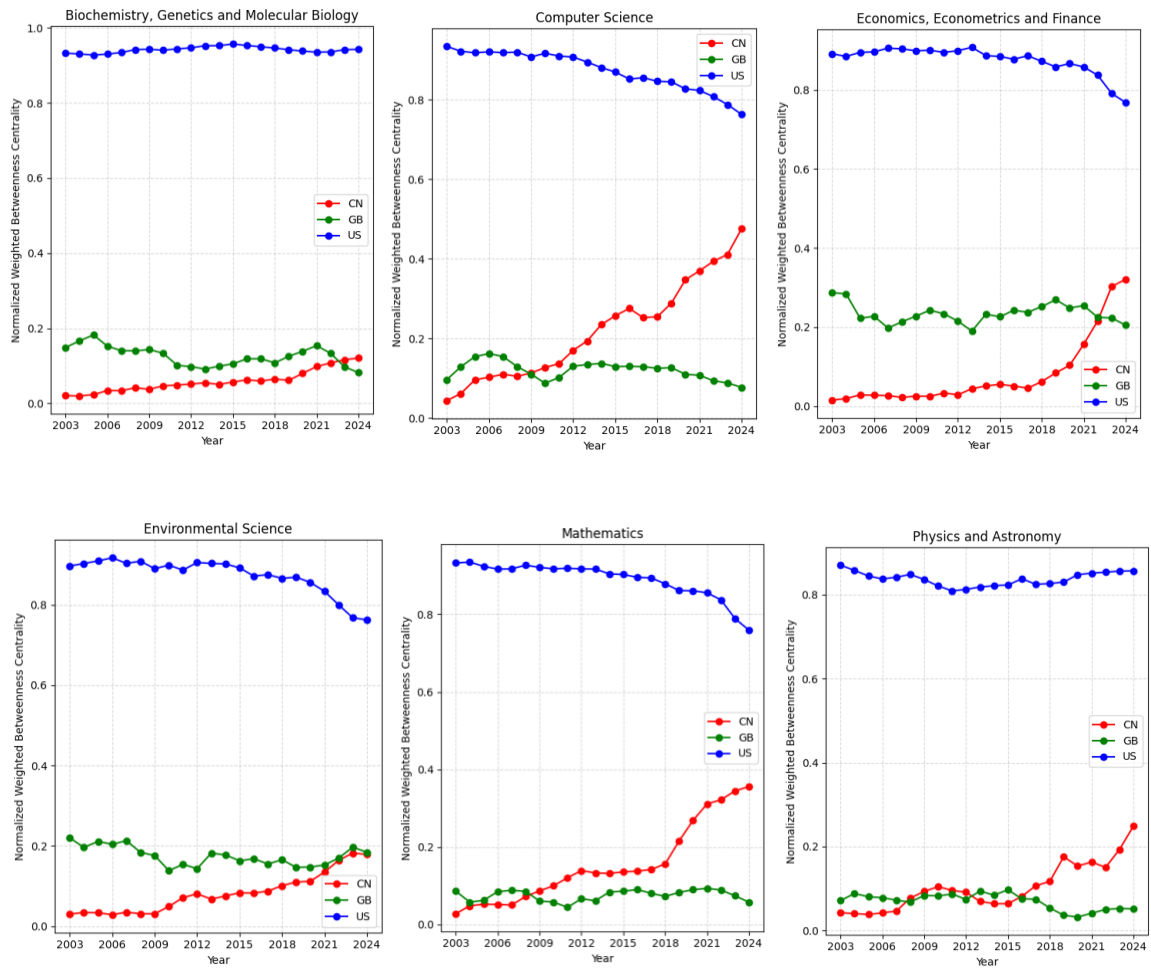